\documentclass[two column,prl]{revtex4}
\usepackage{bm}
\usepackage{graphics}
\usepackage{bm}
\usepackage{graphics}
\usepackage{graphicx}
\usepackage{amsfonts}
\usepackage{amsmath}
\usepackage{amssymb}
\usepackage{graphicx}
\usepackage{color}
\usepackage{fancyhdr}

\begin{document}

\title{Electron Emission Area Depends on Electric Field and Unveils Field Emission Properties in Nanodiamond Films}

\author{Oksana Chubenko$^{1,2}$}\email{chubenko@gwu.edu}
\author{Stanislav S. Baturin$^3$}
\author{Kiran Kumar Kovi$^2$}
\author{Anirudha V. Sumant$^4$}
\author{Sergey V. Baryshev$^2$\vspace{2ex}}\email{sergey.v.baryshev@gmail.com}
\affiliation{$^1$Department of Physics, The George Washington
University, 725 21st St. NW, Washington, DC 20052, USA
\\
$^2$Euclid TechLabs, 365 Remington Blvd., Bolingbrook, IL 60440,
USA
\\
$^3$PSD Enrico Fermi Institute, The University of Chicago, 5640 S.
Ellis Ave, Chicago, IL 60637, USA
\\
$^4$Center for Nanoscale Materials, 9700 S. Cass Ave., Argonne, IL
60439, USA}

\begin{abstract}
In this paper we study the effect of actual, locally resolved,
field emission (FE) area on electron emission characteristics of
uniform semimetallic nitrogen-incorporated ultrananocrystalline
diamond ((N)UNCD) field emitters. To obtain the actual FE area,
imaging experiments were carried out in a vacuum system in a
parallel-plate configuration with a specialty anode phosphor
screen. Electron emission micrographs were taken concurrently with
$I$-$V$ characteristics measurements. It was found that in uniform
(N)UNCD films the field emitting site distribution is not uniform
across the surface, and that the actual FE area depends on the
applied electric field.

To quantify the actual FE area dependence on the applied electric
field, a novel automated image processing algorithm was developed.
The algorithm processes extensive imaging datasets and calculates
emission area per image. By doing so, it was determined that the
emitting area was always significantly smaller than the FE cathode
surface area of 0.152 cm$^2$ available. Namely, the actual FE area
would change from $5\times10^{-3}$ \% to 1.5 \% of the total
cathode area with the applied electric field increased.

We also found that (N)UNCD samples deposited on stainless steel
with molybdenum and nickel buffer layers always had better
emission properties with the turn-on electric field $<$5 V/$\mu$m
and $\beta$-factor of about 1,000, as compared to those deposited
directly onto tungsten having the turn-on field $>$10 V/$\mu$m and
$\beta$-factor of about 200. It was concluded that rough or
structured surface, either on the macro- or micro- scale, is not a
prerequisite for good FE properties. Raman spectroscopy suggested
that increased amount of the graphitic $sp^2$ phase, manifested as
reduced D/G peak ratio, was responsible for improved emission
characteristics.

Finally and most importantly, it was shown that when $I$-$E$
curves as measured in the experiment were normalized by the
field-dependent emission area, the resulting $j$-$E$ curves
demonstrated a strong kink and significant deviation from
Fowler-Nordheim (FN) law, and eventually saturated at a current
density of $\sim$100 mA/cm$^2$. This value was nearly identical
for all studied (N)UNCD films, regardless of the substrate.

\end{abstract}

\maketitle

\pagenumbering{gobble}

\section{I. Introduction}

In general, the electron field emission (FE) properties and thus
efficiency of field emitters are evaluated by plotting the current
density  as a function of the applied electric field $E$ (linear
$j$-$E$ plot representation). For the parallel-plate electrodes
configuration, $E$ is simply a product of the applied voltage $V$
over inter-electrode gap $d$. $j$ is a product of $I$ over $S$,
where $I$ is the current measured in experiment and $S$ is the
emitting surface area. The experimentally determined current
density $j$ is compared to the current density predicted by the
Fowler-Nordheim (FN) law that can be written in a simplified form
proposed by Millikan $j(E) \sim E^2 \exp (-k/E)$, with $k \propto
\frac{\phi^{3/2}}{\beta}$, and $\phi$ being the work function and
$\beta$ being the field enhancement factor, so called
$\beta$-factor. The FN law was originally developed to describe FE
from an ideal flat metallic surface in an ultrahigh electric field
$\sim$ 1 GV/m at 0 K \cite{1}. In many instances, it is convenient
to present properties of field emitters in $ln(j/E^2)$ vs. $1/E$
coordinates (so-called FN plot representation). This helps
understand whether a field emitter under study obeys the FN law,
as well as to calculate the $\beta$-factor from the linear slope
$k$. As seen, both $j$-$E$ plot and FN plot involve the current
density rather than the apparent, as measured, current, which
makes the emission area an important parameter to characterize FE
properties of materials.

Conventionally, the normalization of the apparent current $I$ by
the emission area $S$ is done by using the entire surface area of
a field emission cathode exposed to the electric field (cathode
smaller than anode) or by the entire size of an anode collecting
current in the middle of a cathode (anode is smaller than cathode)
\cite{2}. This has been the standard approach for parallel-plate
configuration FE measurement systems. In either of the ways of, a
small cathode with the area $S$ facing a large anode or a small
anode with the area $S$ collecting electrons from a part of a
large cathode, $S$ is always assumed to be constant. Having $S$
constant is in contrast to locally-resolved FE characteristics of
arrays and patterned/structured surfaces of semiconductors (e.g.
Si and GaN) vastly reported by the groups working collaboratively
at the University of Wuppertal and the Regensburg University of
Applied Sciences \cite{3,4}. In those experiments, a micron-size
anode was scanned across areas of interest with the gap kept
constant. At each location, the applied voltage was increased or
decreased (depending on the location emissivity) in order to set
the emitted current to 1 nA, and 3D maps ($x$-coordinate,
$y$-coordinate, $V$) were recorded. The main result was that $V$
can vary by a factor of 2 to 3 across the area of interest. These
findings suggested that the emission area $S$ had to change with
$V$ in experiments when the micron-size anode would be replaced
with a large plate electrode. This situation can also be referred
to as a problem of non-uniform (strong and weak) emitters.

Eventually, the convention of using the constant emission area:
(i) makes it difficult not only to compare emission
characteristics between different field emitters, but even between
those of the same sample when such a sample undergoes various
treatments or modifications and (ii) may lead to inadequate
interpretation of experimental results. Attempts to avoid the use
of $S$ and representation of field emission data by plotting the
as-measured current $I$ against applied voltage is also a
cumbersome approach as it has no unification between diverse
experimental setups.

Similar to the mentioned locally resolved emission properties of
Si and GaN emitters \cite{3,4}, there were a few reports on
non-uniform emission from carbonic materials such as carbon
nanotubes (CNT) \cite{5,6} and synthetic nanodiamond films
\cite{7}. Emphasis is given to these materials as CNT and
synthetic polycrystalline diamond have long been acknowledged as
promising FE electron sources \cite{8}. They are efficient and
simple to synthesize and scale. While exceptionally high
efficiency of CNT is largely a consequence of their exceptionally
high aspect ratios, nitrogen-incorporated ultrananocrystalline
diamond ((N)UNCD), a highly conductive type of
nanodiamond\cite{9}, is an unconventional field emitter that
performs simply in planar thin film configuration and has turn-on
fields $\sim$10 V/$\mu$m, far below breakdown threshold for any
material. Nonetheless, in Refs. \cite{5,6,7} no attempts to
quantify the actual FE area and to establish the dependence of the
FE area on the applied electric field (if any) have been made. In
this paper, we describe a novel approach to measure FE site
distribution, laterally-resolved on the cathode surface, and to
quantify it by obtaining the dependence of $S$ on the electric
field. This approach was applied to planar thin film (N)UNCD field
emitters grown on stainless steel and tungsten.

\begin {figure}[h]
\includegraphics[width=7cm]{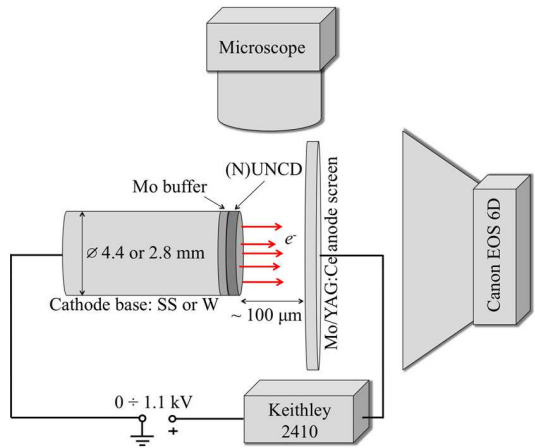}
\caption{Measurement setup diagram (not to scale).} \label{fig:1}
\end {figure}

\section{II. Experiment}

The experiments were performed in a custom imager in the
parallel-plate configuration as described in Ref.\cite{10}. For
convenience, the measurement setup is shown in Fig.1. The imager
has a specialty anode screen which is an optically polished (1
inch dia. and 100 $\mu$m thick) disk made of yttrium aluminum
garnet doped with cerium (YAG:Ce) coated with a Mo film of about
7-8 nm in thickness. The distance between the sample and the anode
screen is set using a micrometer holding the sample. Top and side
view cameras outside the vacuum (side view camera is not depicted
in Fig.1) are used to check the parallelism between the cathode
and anode, and to measure the gap. A Canon DLSR camera is
installed at a viewport behind the anode screen to take pictures
of electron emission patterns, which are formed via
cathodoluminescence (YAG:Ce has 550 nm luminescence line) when
electrons, emitted from the sample and accelerated to the energy
equal to the voltage applied between the electrodes, strike the
YAG:Ce anode screen. Patterns of cathodoluminescence formed on the
YAG:Ce anode screen placed $\sim$100 $\mu$m away from the cathode
represent field electron emission from the sample. The sample
electrode is at ground and the anode frame is isolated and
positively biased. The bias and current readings are enabled by a
Keithley 2410 electrometer. Dwell time at each point to acquire
simultaneously the current and voltage values with their errors,
the vacuum pressure reading and a micrograph is about 5 seconds.
The voltage is swept up and down with a step 1 V from 0 to 1.1 kV.
Therefore, the total time per fully automated experiment is
approximately 3 hours.

To look at the entire cathode surface area, the cathodes were
purposely made smaller than the imaging Mo/YAG:Ce anode screen
which is 1 inch in diameter. The anodes used were 316 stainless
steel (SS) cylinder samples 4.4 mm in diameter, and tungsten
cylinder sample 2.8 mm in diameter. The SS substrate cylinders
were optically polished and the W cylinder substrate had
macroscopic roughness (Fig.2a).

(N)UNCD films were grown using a standard procedure which was
established in our previous studies using a microwave-assisted
chemical vapor deposition system in a mixture of CH$_4$/Ar/N$_2$
with small addition of H$_2$ for initial plasma ignition
\cite{11,12}. To grow (N)UNCD on the SS substrates, Mo buffer of
approximately 110 nm was deposited on SS by magnetron sputtering.
Base pressure in the magnetron system was $<5\times10^{-7}$ Torr.
Prior to coating, the SS cylinders were cleaned \emph{in situ}
using RF discharge plasma. Without breaking vacuum, immediately
after the cleaning, the Mo coating was deposited. Ar was used as a
working gas for both cleaning and sputtering at a pressure of
$\sim$$10^{-3}$ Torr. In addition, to vary surface morphology one
Mo/SS was finished with $\sim$10 nm Ni film that was deposited in
the same magnetron system. We observed that under the same growth
conditions, Ni induced the (N)UNCD film to consist of densely
packed spheres (Fig.2b).

\begin{figure}[h]
\includegraphics[height=3.5cm]{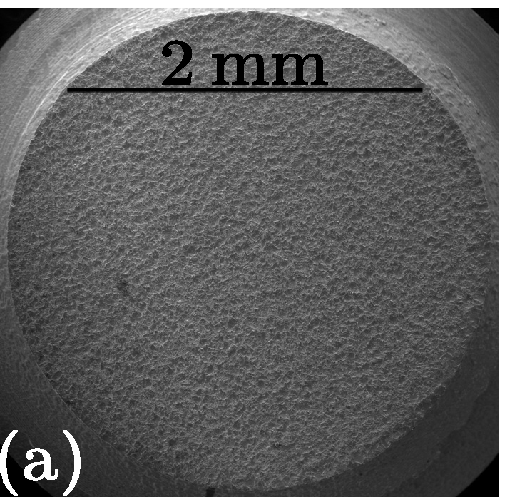} \includegraphics[height=3.5cm]{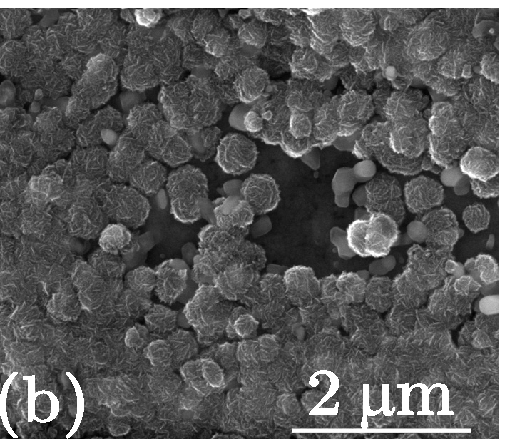}

\

\includegraphics[height=3.5cm]{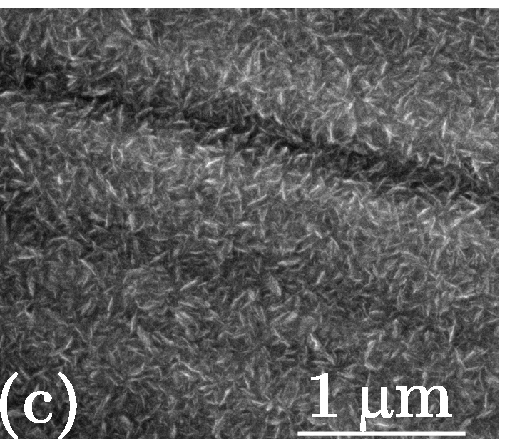} \includegraphics[height=3.5cm]{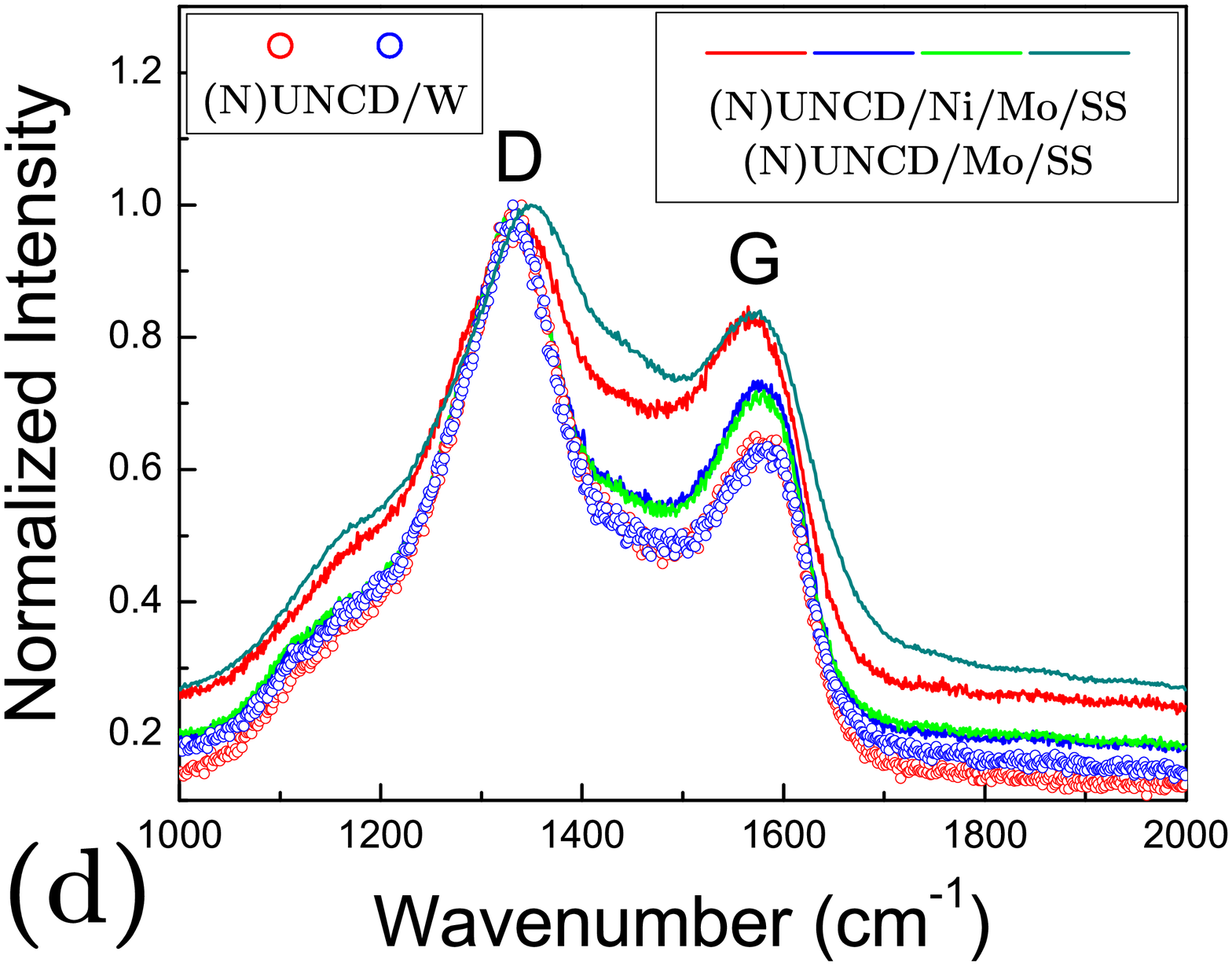}

\caption{SEM micrographs of (a) the original macroscopic surface
roughness of the W substrate; (b) the final morphology of the
(N)UNCD/Ni/Mo/SS sample; (c) needle-like (N)UNCD topography,
identical to all (N)UNCD films regardless of the substrate or
buffer layers; (d) Raman spectra of the samples, showing high
variation (between the samples and across the surface of the same
sample) in the D/G ratio.} \label{fig:2}
\end {figure}

All (N)UNCD films had nearly identical topography as seen by
scanning electron microscopy (SEM), i.e. they had needlelike
nanostructure as illustrated in Fig.2c. Raman spectroscopy
performed using a He-Ne laser ($\lambda$=633 nm) on the other hand
revealed difference between the films on different substrates.
From Fig.2d, it follows that the D/G ratio was 1.2 to 1.4 for the
(N)UNCD on the SS substrates while the D/G was 1.6 for the (N)UNCD
on tungsten. Smaller D/G ratio suggests higher content of
graphitic $sp^2$ phase in (N)UNCD. This behavior is consistent to
the previous studies as reported in Refs.\cite{13,14,15}. This is
also consistent with our previous measurements: an (N)UNCD film on
pure Mo substrate featured a D/G=1.6 \cite{11}, while an (N)UNCD
on Mo/SS had a D/G=1.3 \cite{12}. The resistivity of the (N)UNCD
films is assumed to be 0.1 $\Omega\times$cm, as suggested by a
four-probe measurement of a (N)UNCD film grown on an insulating Si
witness coupon under the same growth conditions.

\section{III. Results and Discussions}

Four datasets obtained from the measurements are presented here:
one from (N)UNCD/Mo/SS emitter, two sets taken at different
inter-electrode gaps (147 and 106 $\mu$m) from (N)UNCD/Ni/Mo/SS
emitter, and one set from (N)UNCD/W emitter (we will further label
the datasets as \textbf{MoSS}, \textbf{NiMoSS1}, \textbf{NiMoSS2},
and \textbf{W}, respectively). Fig.3, showing one image per
dataset, illustrates emission patterns captured by the camera
behind the Mo/YAG:Ce anode screen (see Fig.1) -- the green light
patterns are caused by the process of cathodoluminescence when
field emitted electrons accelerated to a few hundred or a thousand
eV bombard the phosphor. This means the patterns represent field
electron emission from (N)UNCD cathode projected onto the
Mo/YAG:Ce anode screen placed $\sim$100 $\mu$m away from the
cathode. Comparing results presented in Fig.3, our initial
qualitative conclusions were that (i) the field emission site
distribution in topographically uniform (N)UNCD thin films is not
uniform and (ii) the (N)UNCD samples grown on the SS base would
emit larger currents with emission distributed more uniformly as
compared to the (N)UNCD on the W base.

\begin{figure}[h]
\includegraphics[width=7cm]{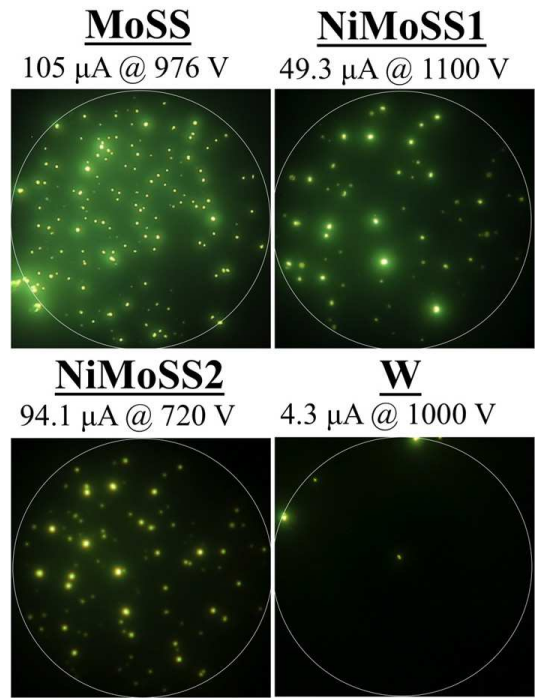}
\caption{Imaging chart of the four imaging datasets obtained from
the FE measurements. The chart compares emissivity of the samples
at the maximal electric field and current.} \label{fig:3}
\end {figure}

To enable accurate quantitative analysis of the extensive FE
micrograph datasets described and discussed below, we made use of
a clustering-based algorithm class. Such algorithms, for instance,
are common tools in quantitative image analyses in astrophysics
\cite{16} and biology \cite{17} used for accurate discrimination
of a true signal from the background. The procedure in its
entirety is explained in Appendix A and the core concept is
mentioned in brief here. The core concept of the method is based
on digitization of FE imaging micrographs and partitioning of
resulting data into groups, or clusters, of similar elements. The
image-clustering algorithms are adapted as required by using the
strategies and approaches as described in \cite{18,19,20}. The
entire analysis procedure is implemented in Wolfram Mathematica to
run in a parallel configuration on a multi-core computing cluster
such that each processor is dedicated to work on its prescribed
image. By doing so, it becomes possible to process about 100
images concurrently; reducing the total computation time
dramatically. This method does not require a priori knowledge
about the number and shape of clusters and can be used as an
automated cathode performance test. The pixel size in cm$^2$ is
referenced to the full image size in pixels and the known diameter
of the (N)UNCD cathode, 0.44 or 0.28 cm. The product of the total
number of pixels combined by the clustering procedure and the
pixel size in cm$^2$ provides the total emission area per
micrograph in the datasets shown in the Appendices B to E. As a
specific example of how an image is converted into a pixel map of
a known field emission area, we use a micrograph from the dataset
\textbf{MoSS}, 21.8 $\mu$A@650V, Appendix B. The width of the
image in pixels is equal to the sample diameter $d$, with $d =$
0.44 cm, which allows for calculating a single pixel area
$S_{pix}$ as small as $9.9\times10^{-7}$ cm$^2$. Multiplication of
the total number of selected pixels by $S_{pix}$ yields the total
emission area of $4\times10^{-4}$ cm$^2$ (Fig.4).

\begin {figure}[h]
\includegraphics[width=5cm]{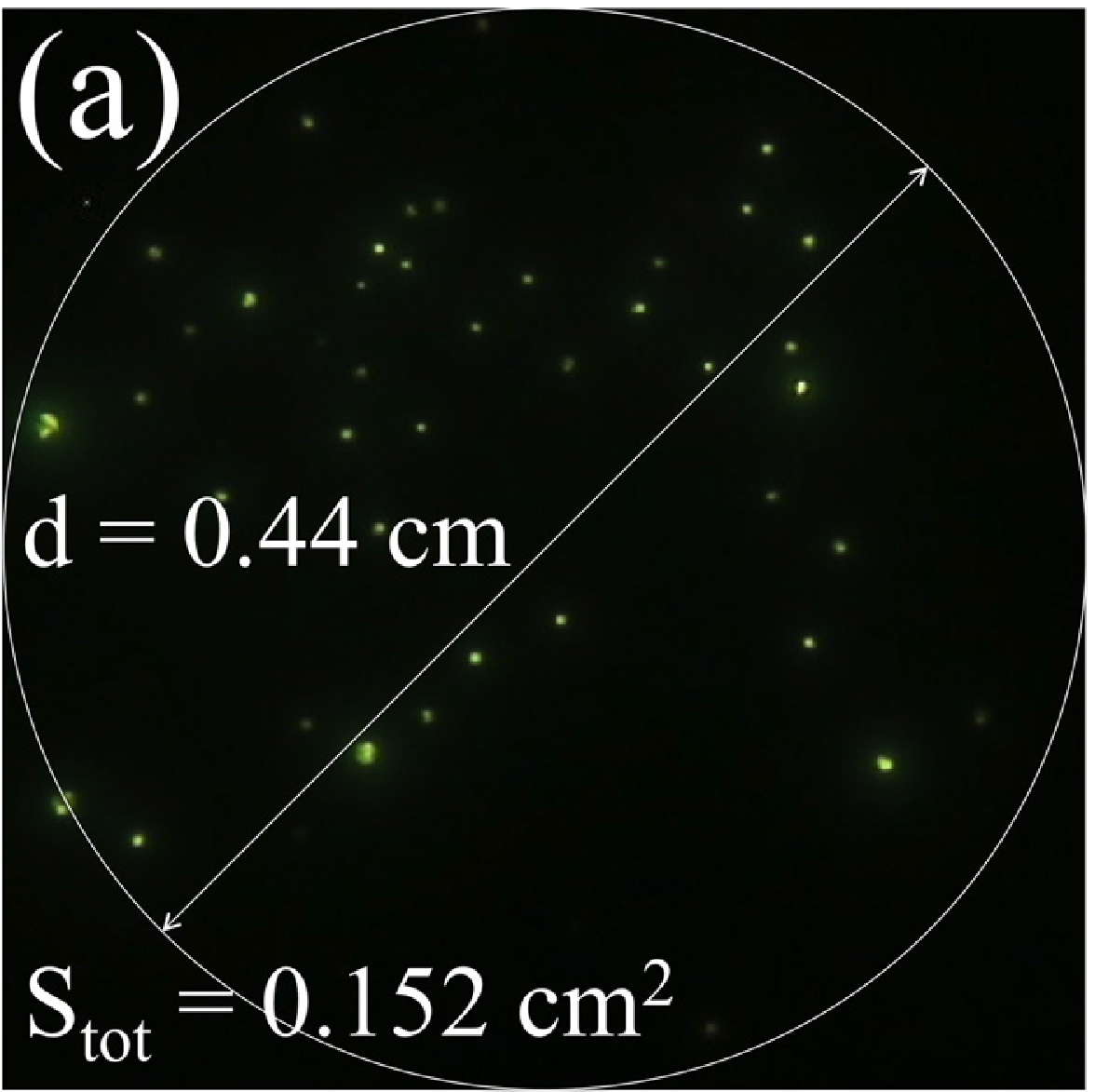}

\

\hspace*{-1.cm} \includegraphics[width=5.9cm]{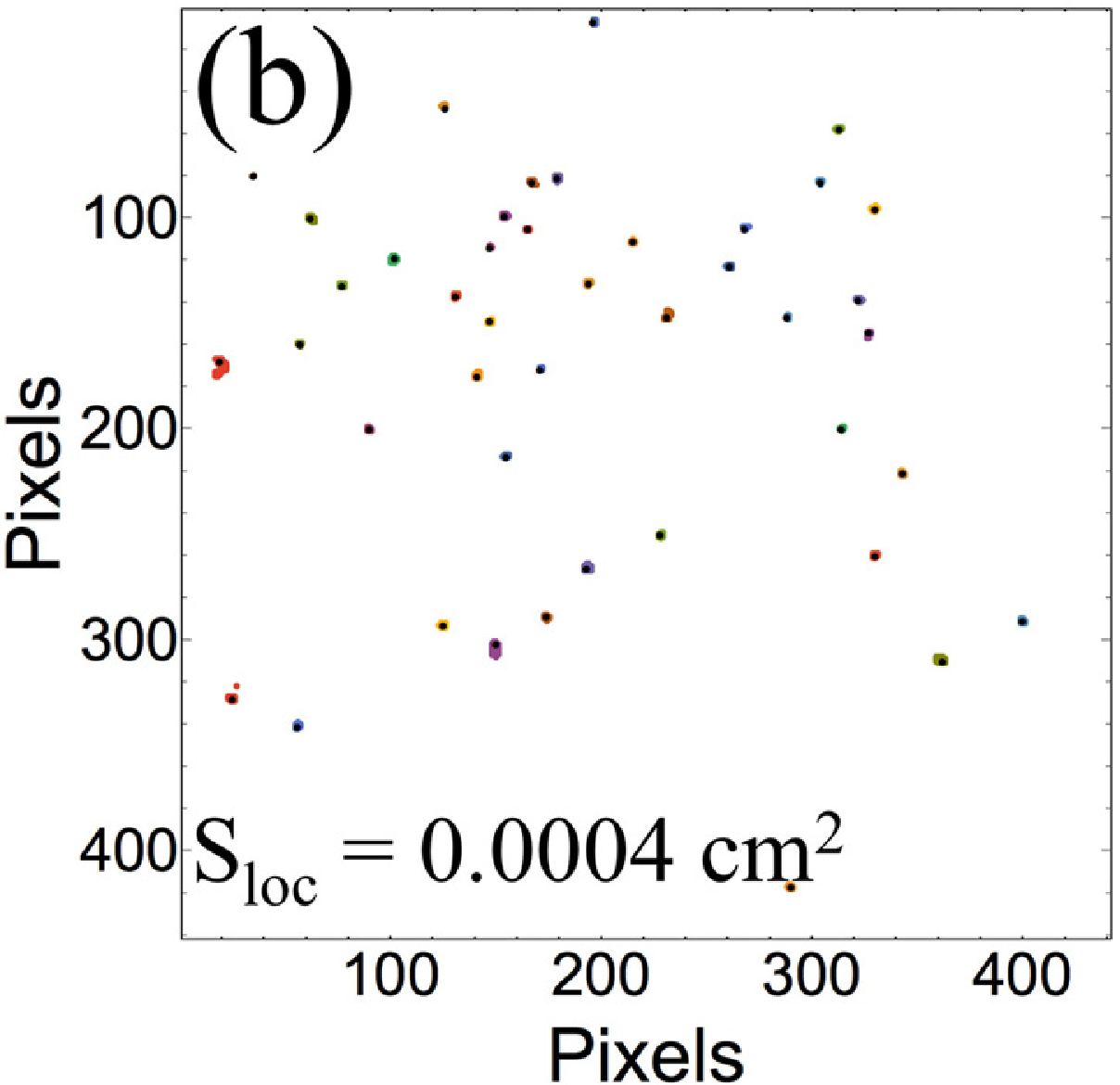}

\caption{(a) One exemplary micrograph of the emission pattern of
the sample (N)UNCD/Mo/SS, seen on the YAG anode screen at 21.8
$\mu$A and 650V, from the dataset \textbf{MoSS}, Appendix B. The
total cathode area is $S_{cathode}=$0.152 cm$^2$. (b) The green
light emission pattern imaged on the YAG anode screen converted
into the actual locally resolved emission area $S_{loc}(E)$ by the
clustering algorithm: the total number of pixels clustered
together is multiplied by the pixel size in cm$^2$, which is
evaluated with respect to the known total cathode area. The
emission area appears to be only $4\times10^{-4}$ cm$^2$ at the
current of 21.8 $\mu$A.} \label{fig:4}
\end {figure}

The set \textbf{MoSS} was measured manually starting at 3.8
V/$\mu$m when reasonably bright YAG:Ce emission was detected with
the camera at ISO=1,000 and a shutter speed of 1 second. After
that, images were taken every 50 V until the maximum current of
100 $\mu$A was reached. Therefore, there are 11
"current-voltage-image" measurements in this case (see Appendix
B).

For other samples/sets, our most recent LabVIEW software was
configured such that the imaging camera took images every 20 V
(sets \textbf{NiMoSS1} and \textbf{NiMoSS2} in Appendices C and D)
and 10 V (set \textbf{W}, Appendix E) as voltage was first ramped
up and then down. $I$-$V$ measurements for all sets
\textbf{NiMoSS1}, \textbf{NiMoSS2} and \textbf{W} were performed
with a step of 1 V. All image datasets (see appendices) were
processed using the automated algorithm, as described is Appendix
A, further to evaluate the actual FE area and determine its
dependence on the electric field, $S_{loc}(E)$.

For all four datasets in discussion, we plotted the current
density $j$ versus the electric field in Fig.5 by convention, i.e.
by dividing the experimentally measured current by the entire
cathode area $S_{cathode}=const$
\begin{align}
\label{eq:1} j(E)=\frac{I(E)}{S_{cathode}}.
\end{align}

\begin{figure*}
\includegraphics[width=7cm]{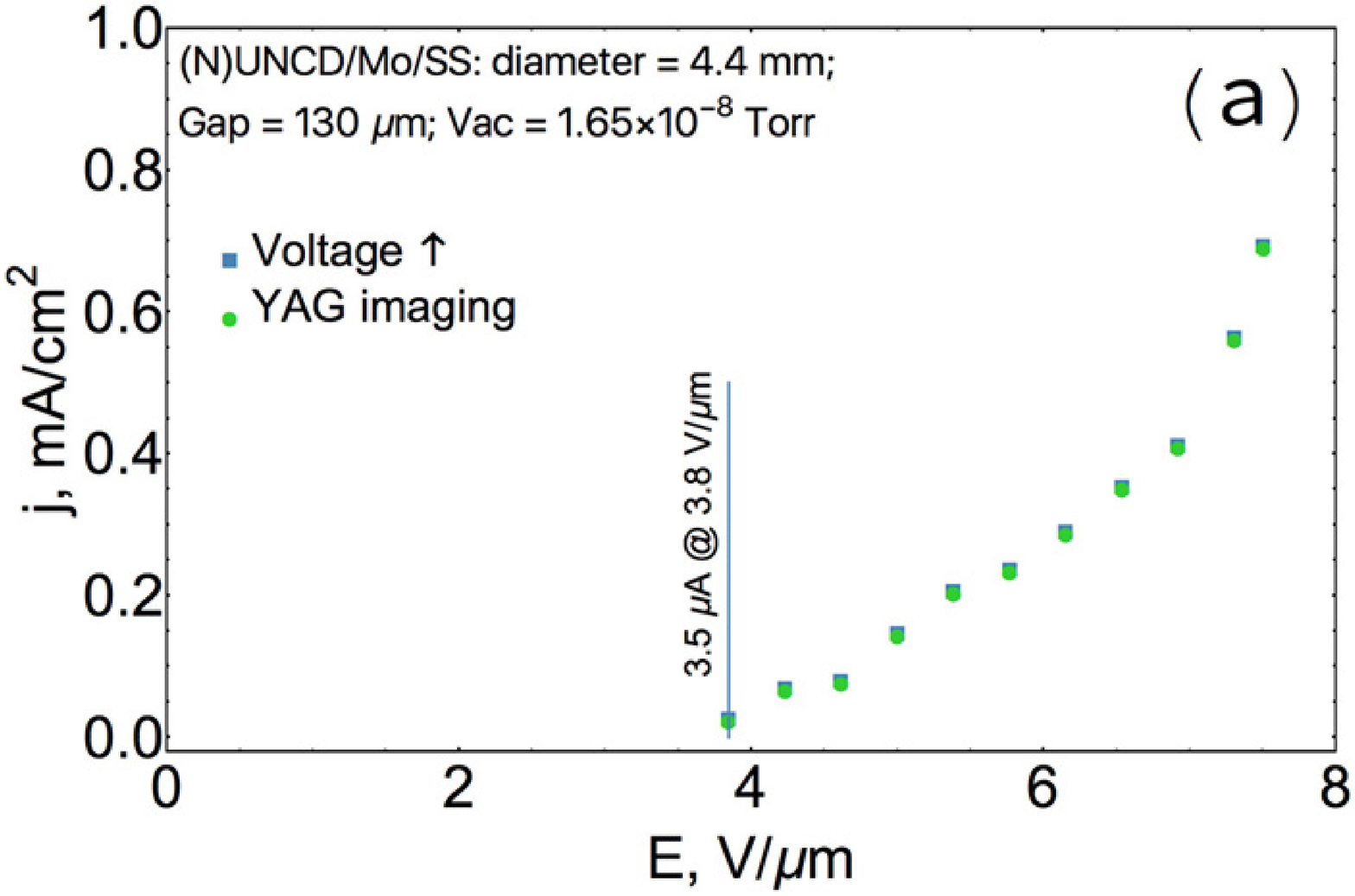}  \includegraphics[width=7cm]{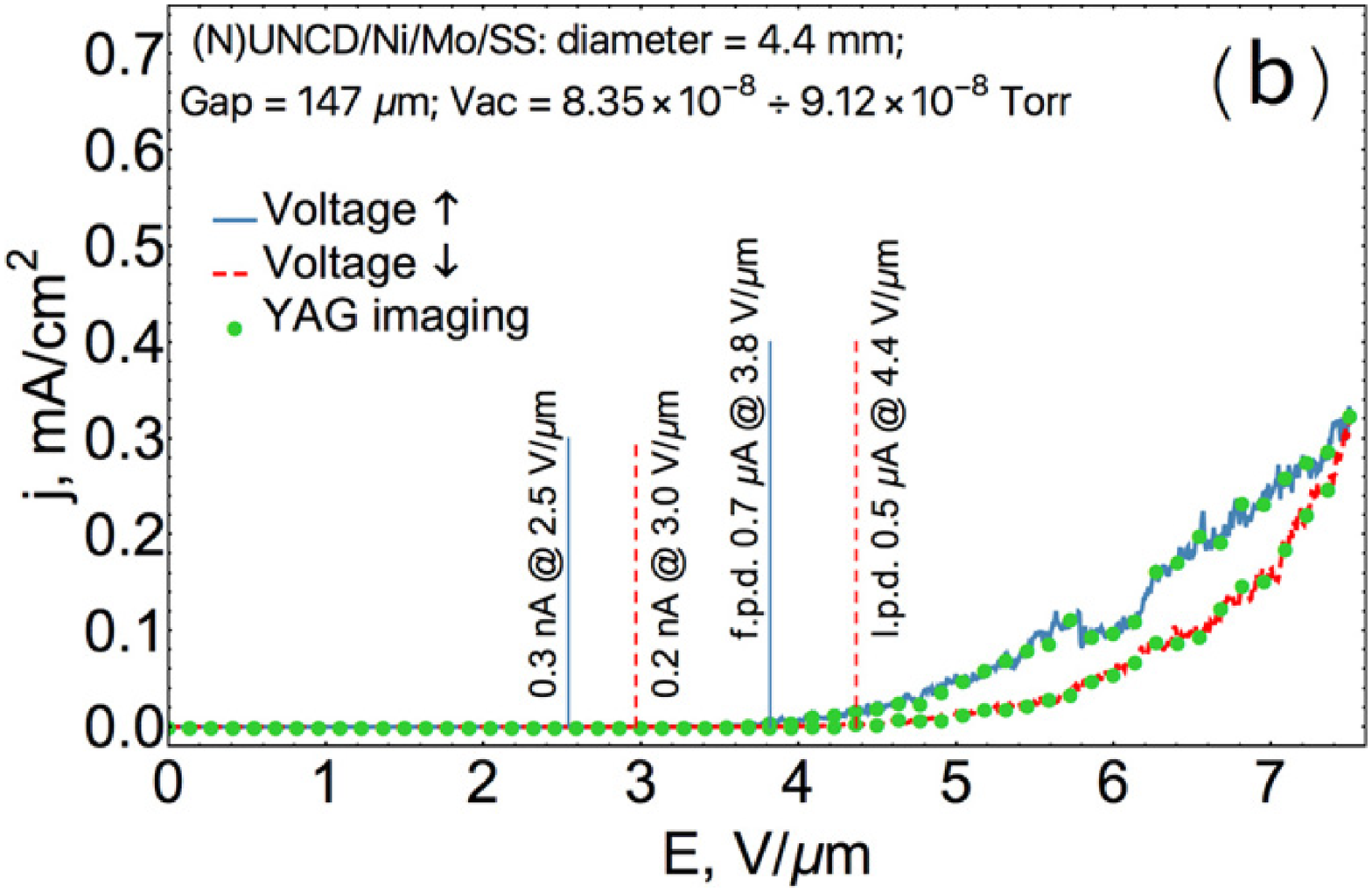}

\includegraphics[width=7cm]{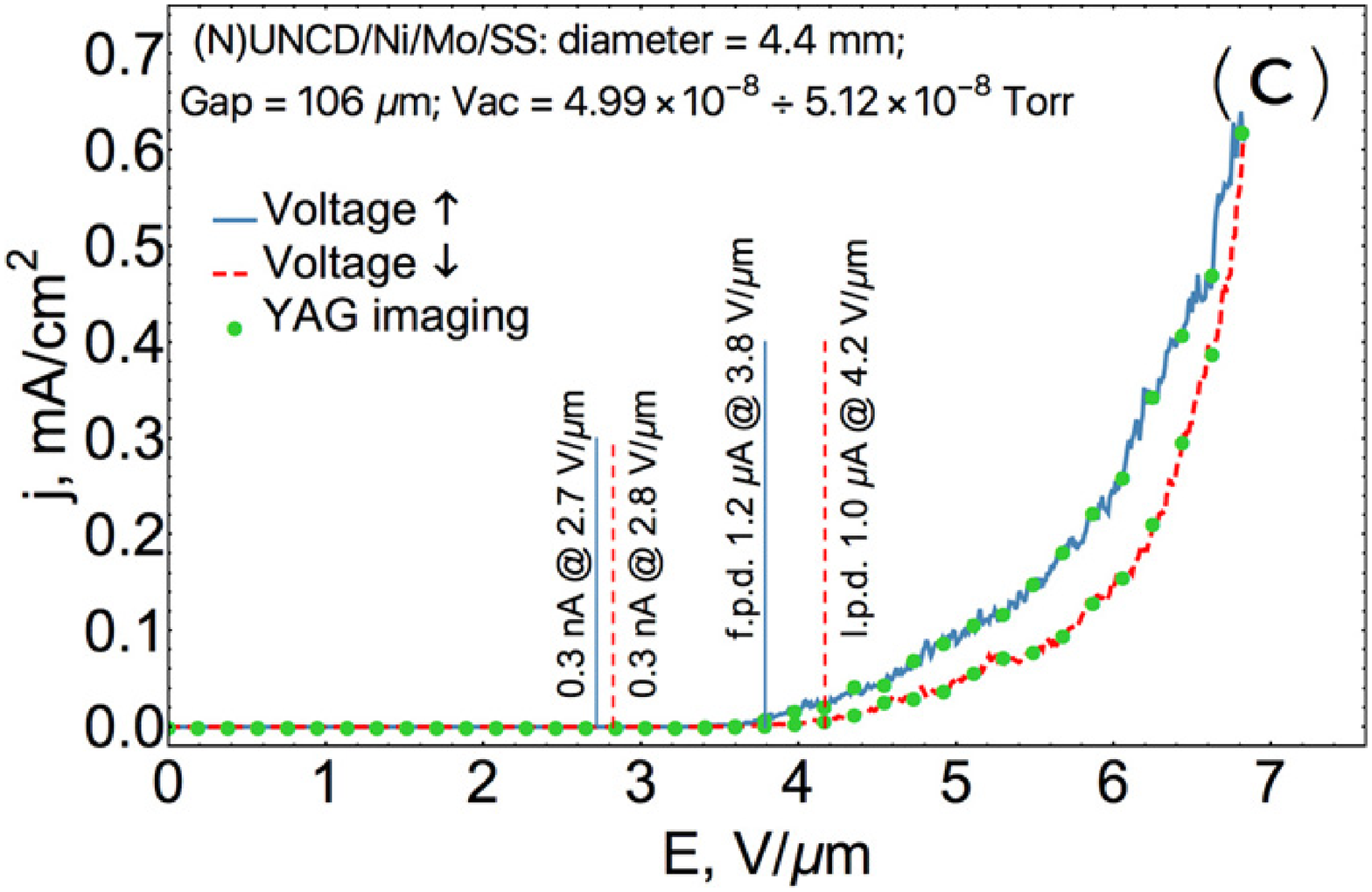} \includegraphics[width=7cm]{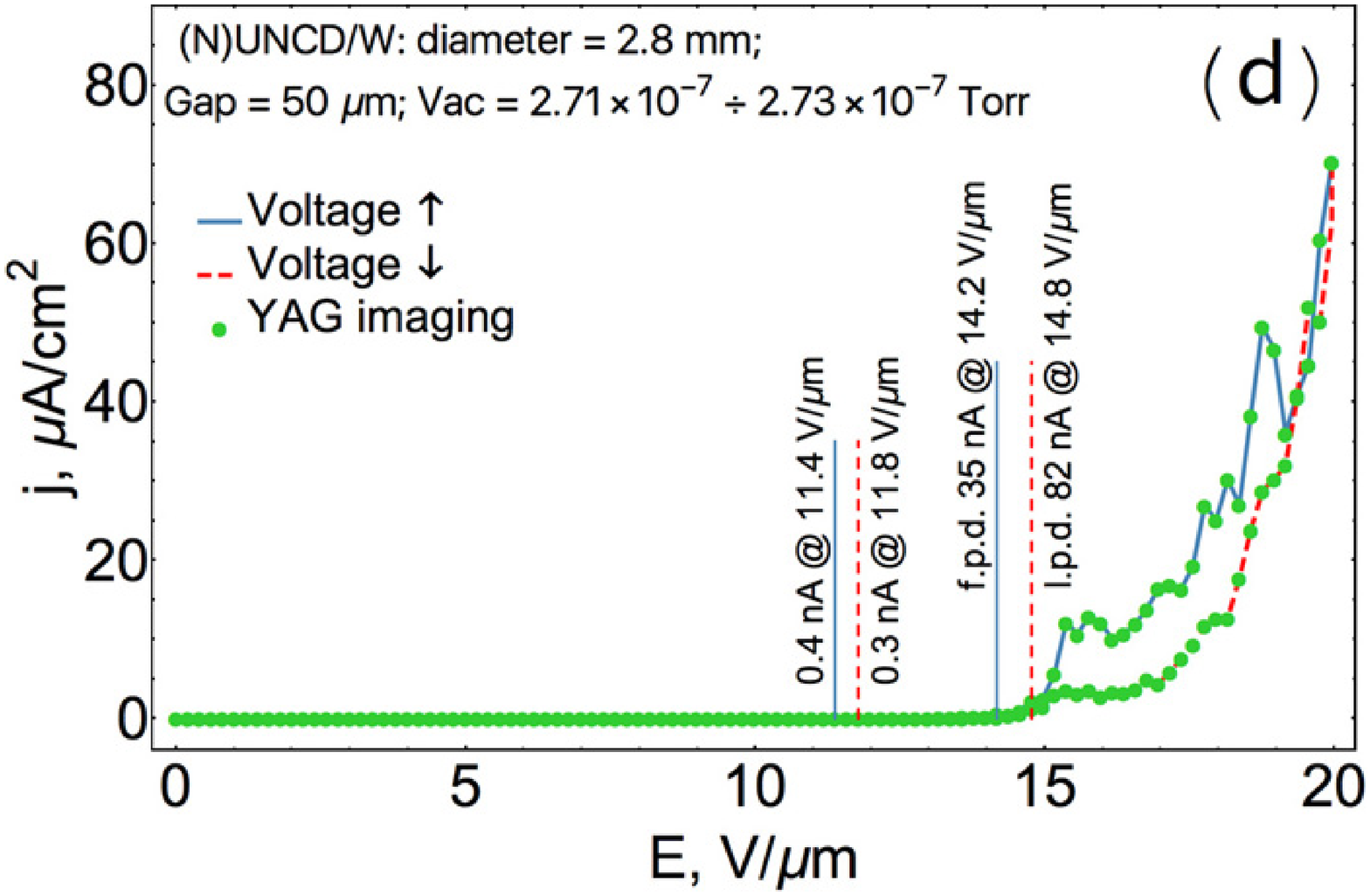}
\caption{$j$-$E$ plots for the datasets (a) \textbf{MoSS}, (b)
\textbf{NiMoSS1}, (c) \textbf{NiMoSS2}, and (d) \textbf{W}.
Current density $j$ is calculated using Eq.1.} \label{fig:5}
\end{figure*}

\begin{figure*}
\includegraphics[width=7cm]{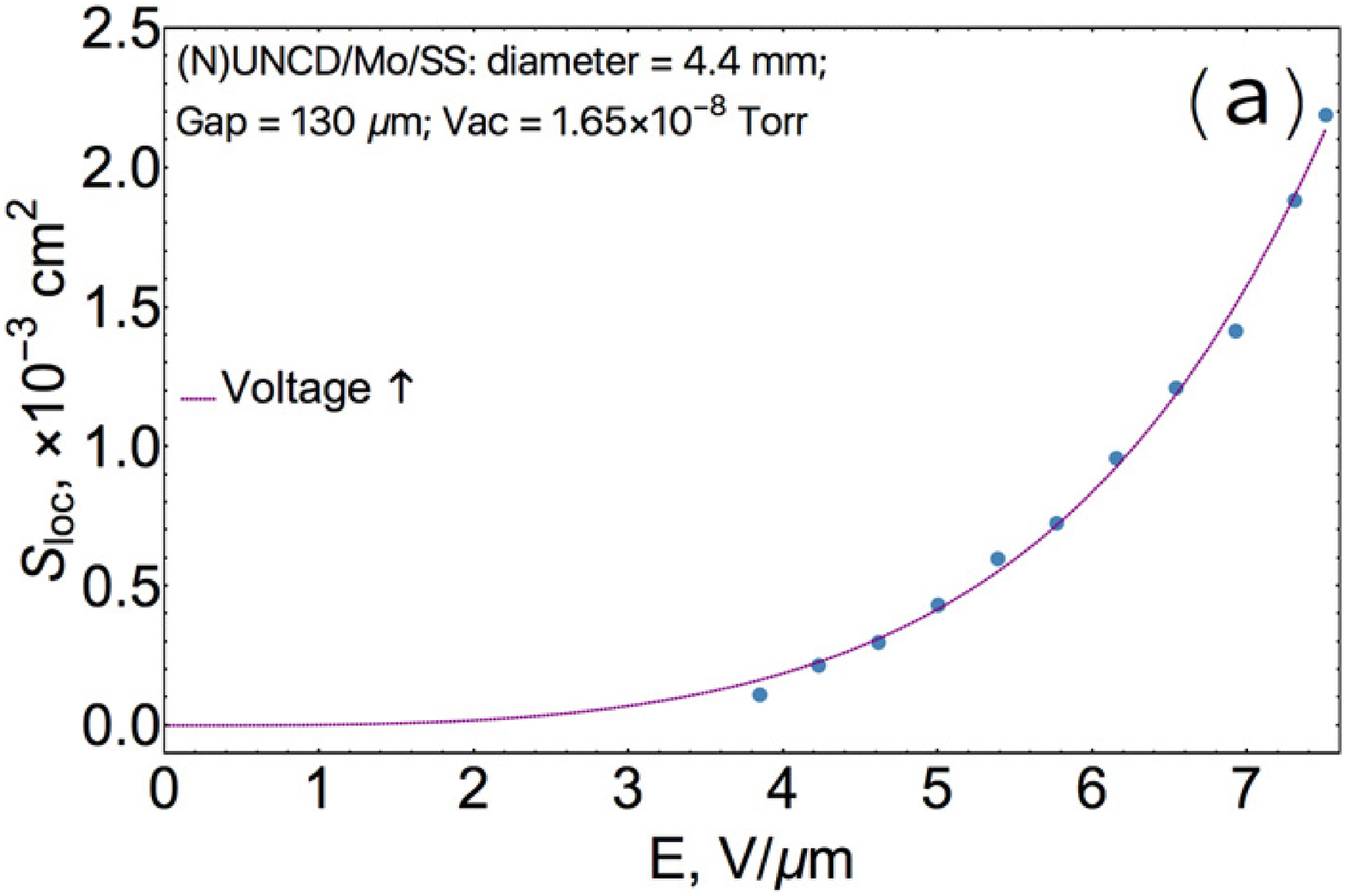} \includegraphics[width=7cm]{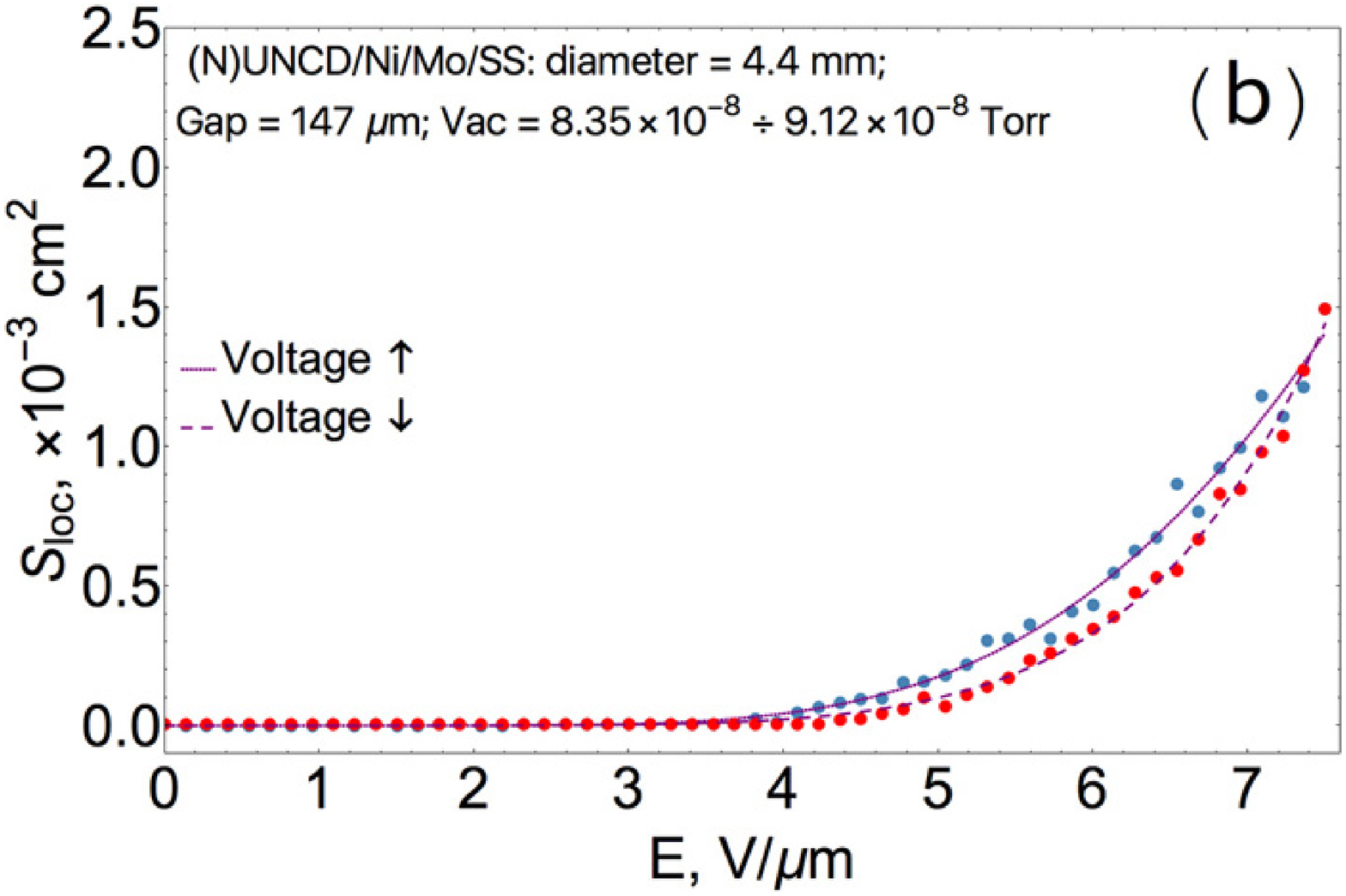}

\includegraphics[width=7cm]{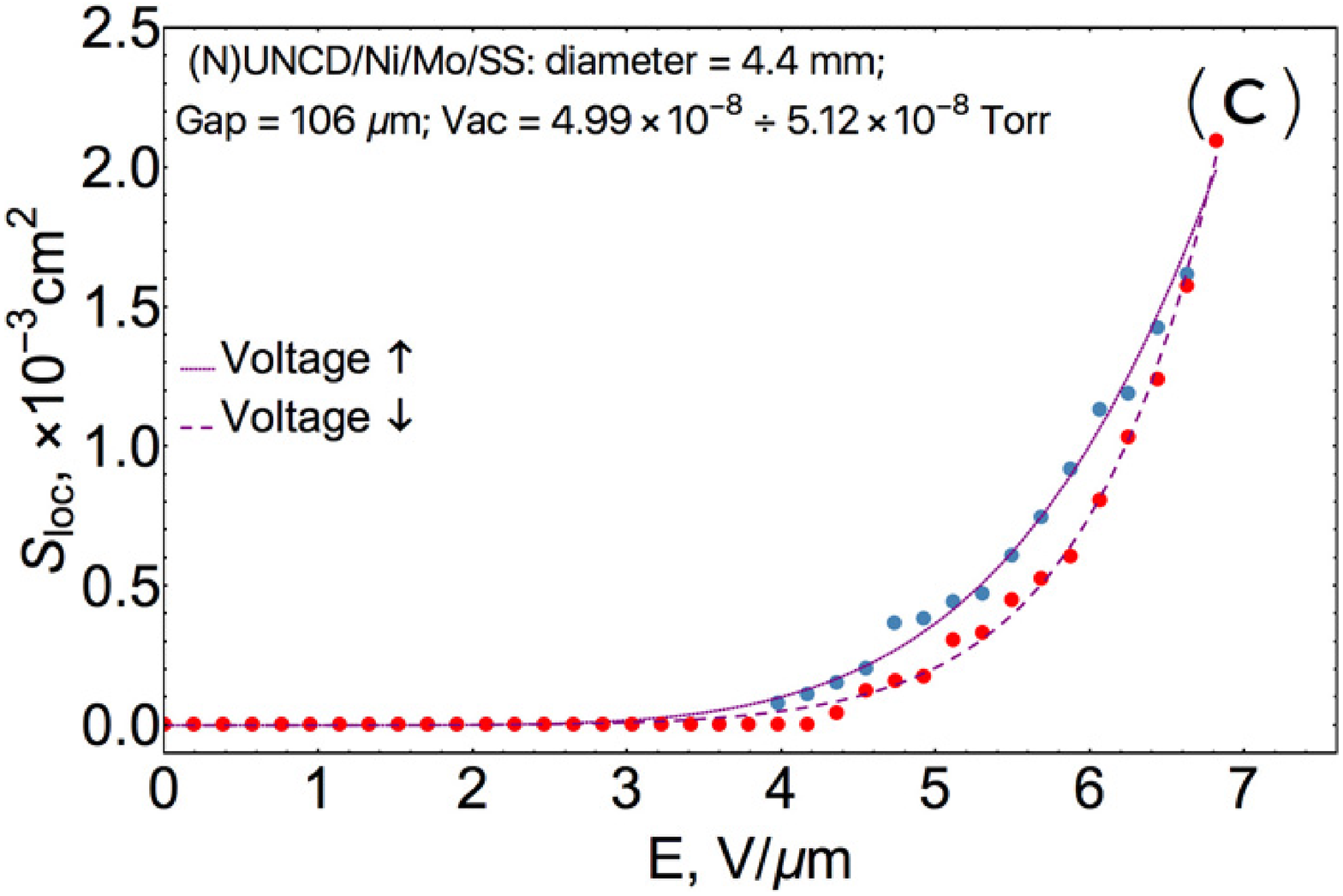} \includegraphics[width=7cm]{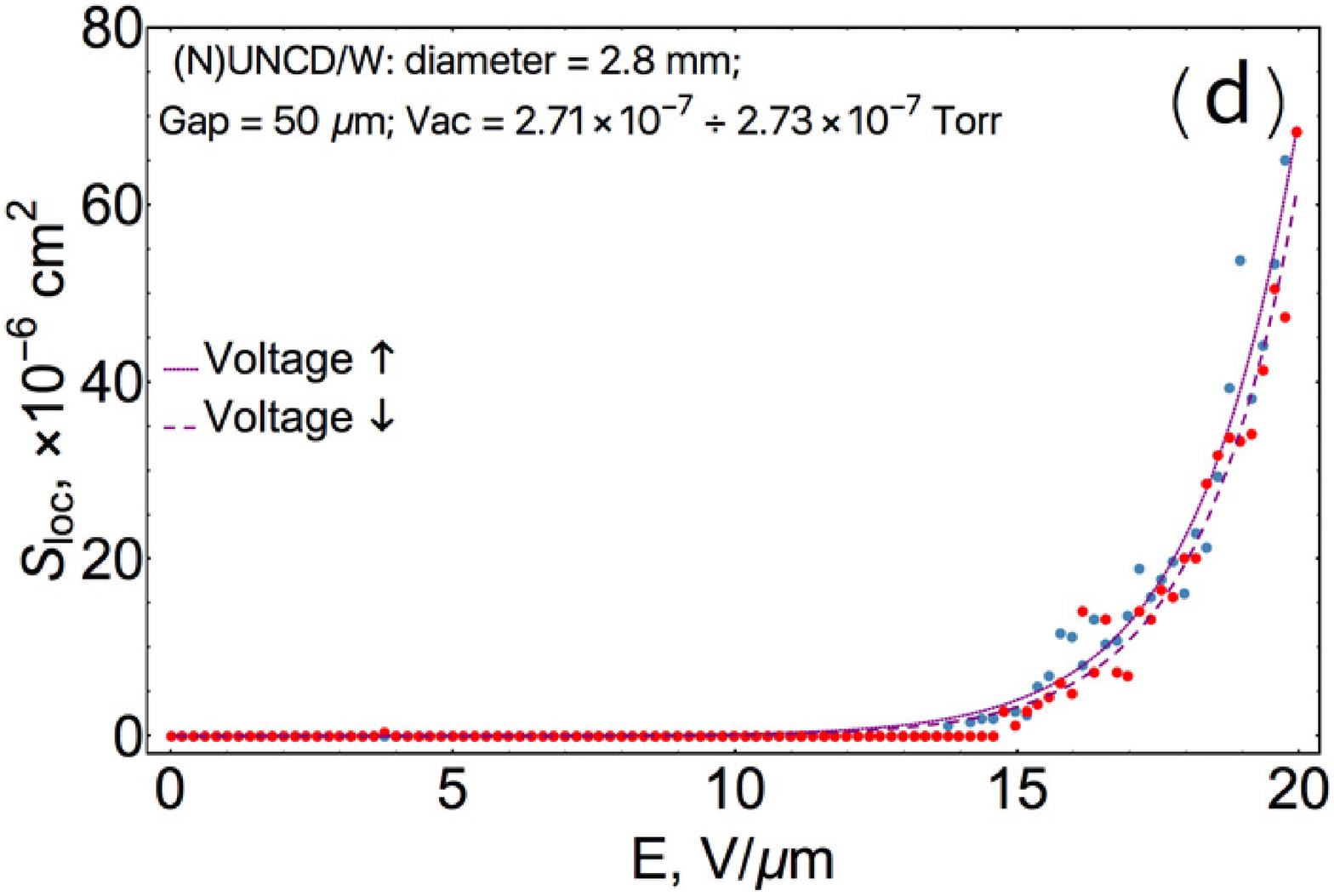}

\caption{Dependence of the actual field emission area $S_{loc}$ on
the electric field for all the datasets: (a) \textbf{MoSS}, (b)
\textbf{NiMoSS1}, (c) \textbf{NiMoSS2}, and (d) \textbf{W}. The
dependencies $S_{loc}(E)$ were obtained using the developed image
processing algorithm.} \label{fig:6}
\end{figure*}

\noindent The conventional representation makes the samples seem
to enable different current densities. This results in somewhat
misleading interpretation of data even for the same emitter
(N)UNCD/Ni/Mo/SS measured at two different gaps (panels (b) and
(c) in Fig.5).

Note, in Figs.5, 7, and 8, the low field vertical markers show the
turn-on electric field, i.e. the electric field at which field
emission initiates. The higher field markers show the current (and
the corresponding electric field) at which the image processing
algorithm has detected the first pixel (first pixel detected,
f.p.d., voltage ramped up) and the last pixel (last pixel
detected, l.p.d., voltage ramped down) with the intensity that
satisfies Eq.A4 in Appendix A. The datasets in the Appendices B to
E start/end with the images before/after which the imaging
processor would not detect bright pixels above the background.

After applying the imaging algorithm to the datasets shown in full
in Appendices B to E, the locally resolved actual emission area
$S_{loc}$ was found to be a dynamic property that strongly depends
on the electric field; for all the (N)UNCD field emitters
measured. $S_{loc}(E)$ non-linearly but monotonically increased
with the applied electric field, as illustrated in Fig.6, as well
as monotonically decreased with the electric field swept down to
zero following nearly identical non-linear law. The samples on the
SS bases had emission area of $\sim$1\% of the total cathode area
(4.4 mm dia.) at the maximum output current of 100 $\mu$A and
electric field of about 7 V/$\mu$m. The emission area of the
sample on W was as small as $\sim 10^{-3}$\% of the total cathode
area (2.8 mm dia.) at the maximum output current of 5 $\mu$A and
electric field of 20 V/$\mu$m.

The apparent inconsistency revealed in Fig.5 vanishes if the
experimentally measured current is normalized by the dynamic
surface area $S_{loc}(E)$, which was obtained by the least square
fitting of the experimental curves using a combined function
$S_{loc}(E)=a E^n \exp(b E^m)$

\begin{align}
\label{eq:2}
j_{cor}(E)=\frac{I(E)}{S_{loc}(E)},
\end{align}
with $S_{loc}(E)\neq const$. The corrected semi-log $j_{cor}$-$E$
plots are summarized in Fig.7, as well as the comparison between
the conventional (left column (a-d)) normalization using
$S_{cathode}$ and the newly proposed (right column (a-d))
normalization using $S_{loc}(E)$ is drawn. From Figs.7 and 8, it
is seen that all the samples demonstrated saturation behavior.
After a very short FN-like dependence manifested as a nearly
linear $j$-$E$ relation (only available for the automatically
recorded datasets \textbf{NiMoSS1}, \textbf{NiMoSS2} and
\textbf{W}), a strong kink and deviation from the FN law took
place. In fact, $j_{cor}$-$E$ still slowly increases as  when
plotted in linear coordinates. The lingo \textit{saturation} is
used because $ln(I/E^2)$ indeed saturates with $E$, as illustrated
in Fig.8, because the current increment is the same or smaller
than that of $E^2$.

Also note, at small voltages applied with no field emission
current, $S_{loc}(E)=0$. In this region, the resulting current
density is infinite. To smoothly stitch the parts of the
$j_{cor}$-$E$ curves before and after the turn-on field, the
currents $I(E)$ in the sets \textbf{NiMoSS1}, \textbf{NiMoSS2},
and \textbf{W} measured between 0 V/$\mu$m and the sub-nA vertical
markers identifying the turn-on field are normalized by
$S_{loc}(E)=2.7\times10^{-6}$ cm$^2$,
$S_{loc}(E)=7.9\times10^{-6}$ cm$^2$, and
$S_{loc}(E)=4.4\times10^{-7}$ cm$^2$, respectively, as obtained by
the fitting $S_{loc}(E)$ at the turn-on field point. This yields
current densities $\sim10^{-2}$ mA/cm$^2$. When $j$ is calculated
by the convention with $S_{cathode}(E)=const$, the measured
current is normalized by the total cathode surface area of 0.152
cm$^2$, or 0.062 cm$^2$ for \textbf{W} dataset. This leads to a
few orders of magnitude difference between $j$ and $j_{cor}$
(Fig.7), and sets $j_{cor}$-$E$ curves higher against $j$-$E$
curves in FN coordinates in Fig.8.

\begin{figure*}
\includegraphics[width=0.9 \textwidth]{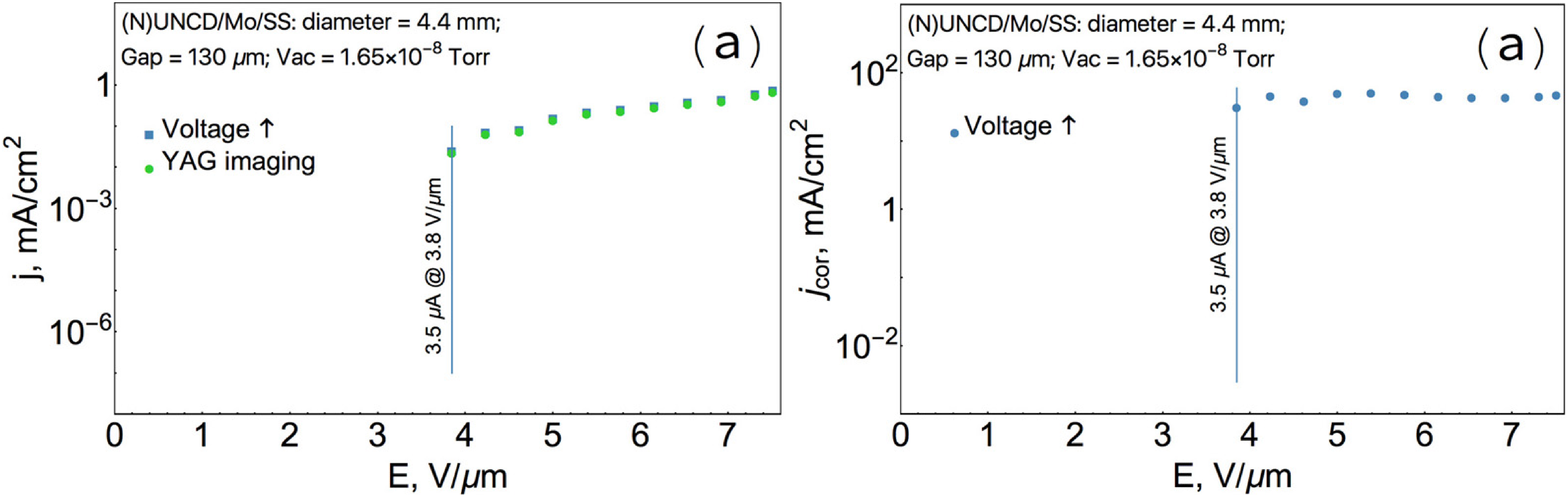}
\includegraphics[width=0.9 \textwidth]{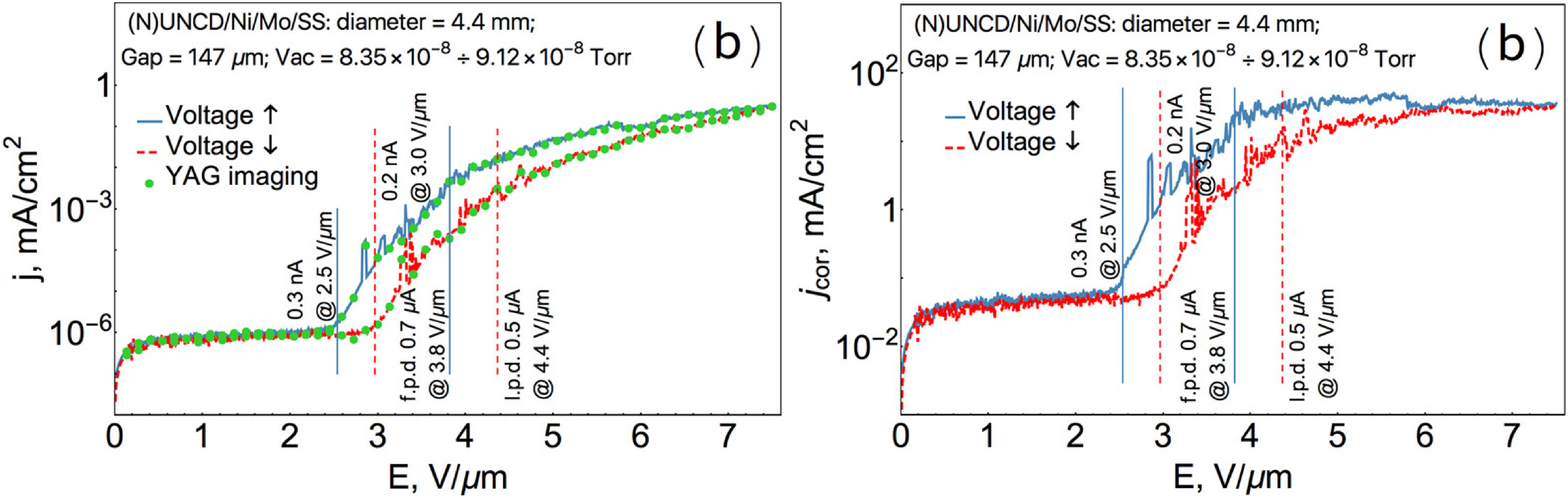}
\includegraphics[width=0.9 \textwidth]{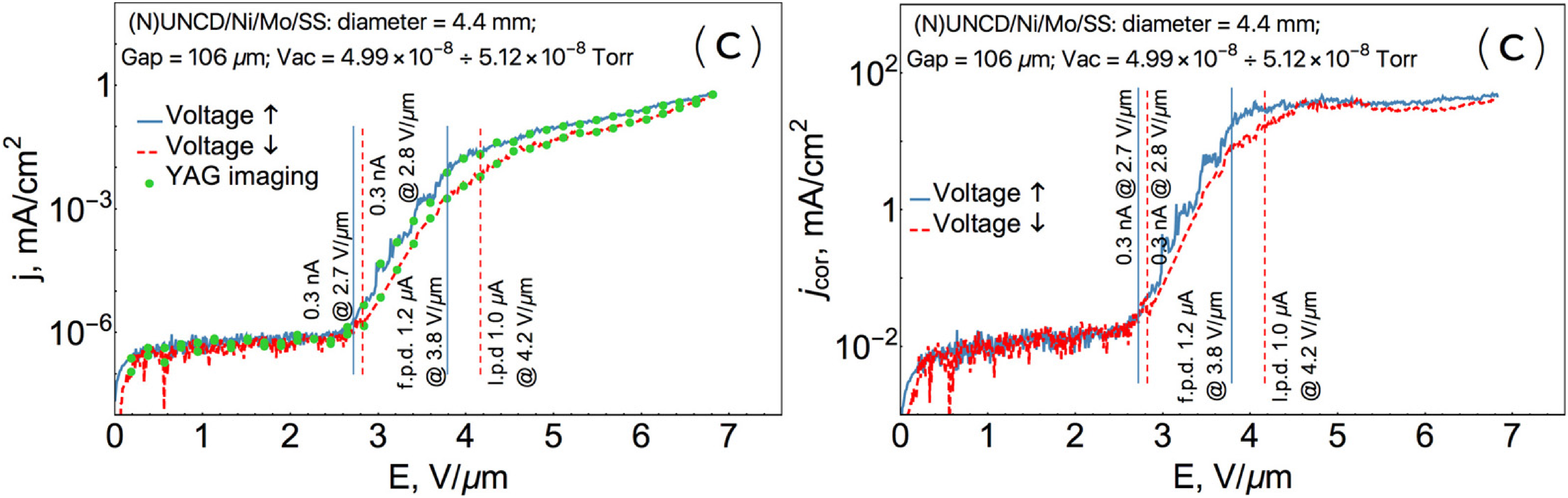}
\includegraphics[width=0.9 \textwidth]{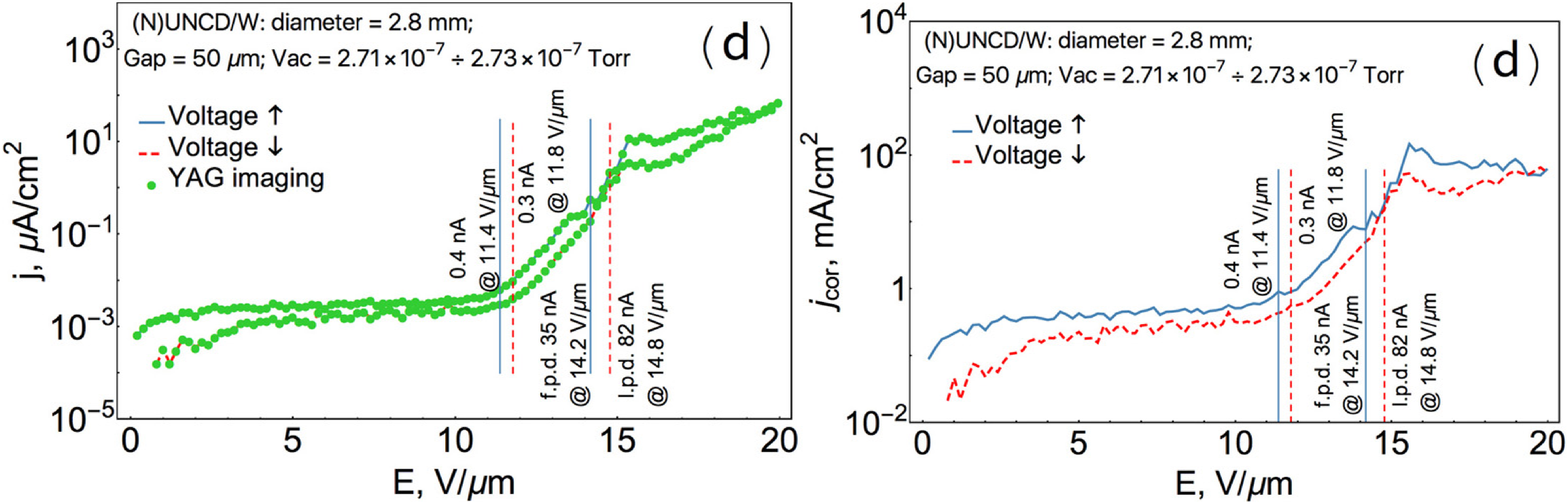}
\caption{Semi-log plots of current density as a function of the
applied electric field obtained by convention using Eq.1 ($j$ vs.
$E$, (a) to (d) on the left) and by normalizing with $S_{loc}(E)$
using Eq.2 ($j_{cor}$ vs. $E$, (a) to (d) on the right) for four
samples: (a) \textbf{MoSS}, (b) \textbf{NiMoSS1}, (c)
\textbf{NiMoSS2}, and (d) \textbf{W}.} \label{fig:7}
\end{figure*}

\begin{figure*}
\includegraphics[width=0.9 \textwidth]{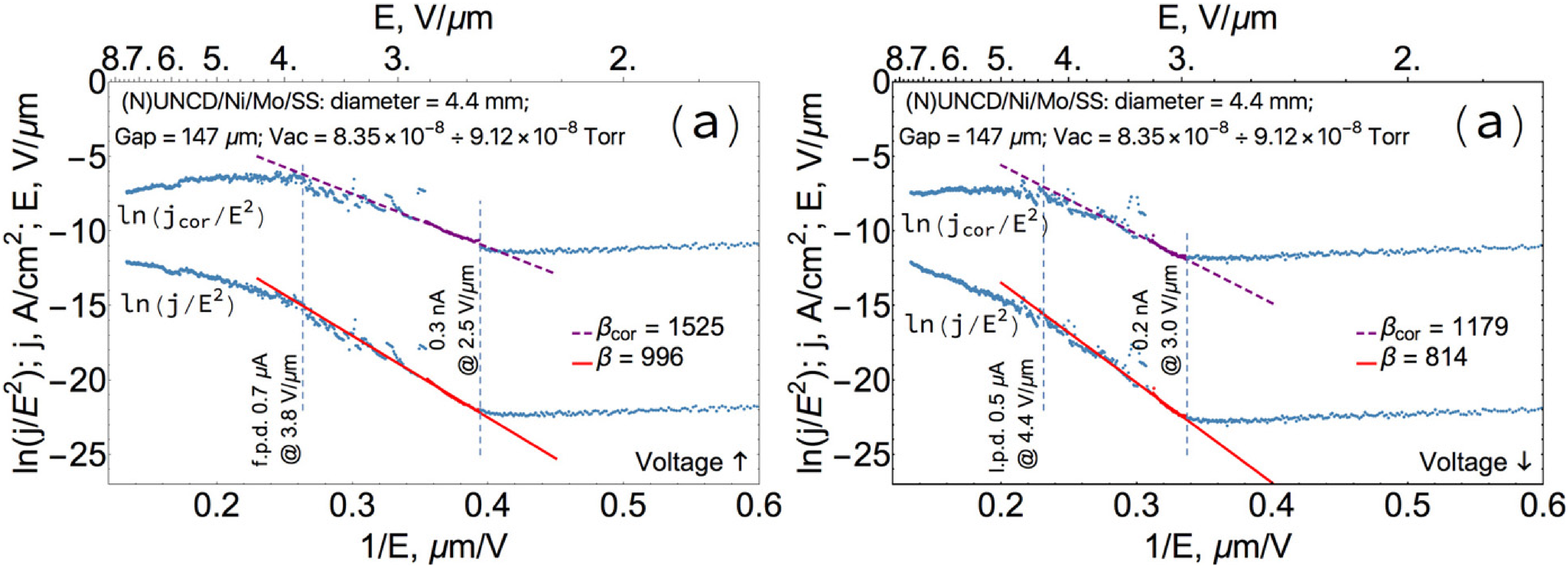}
\includegraphics[width=0.9 \textwidth]{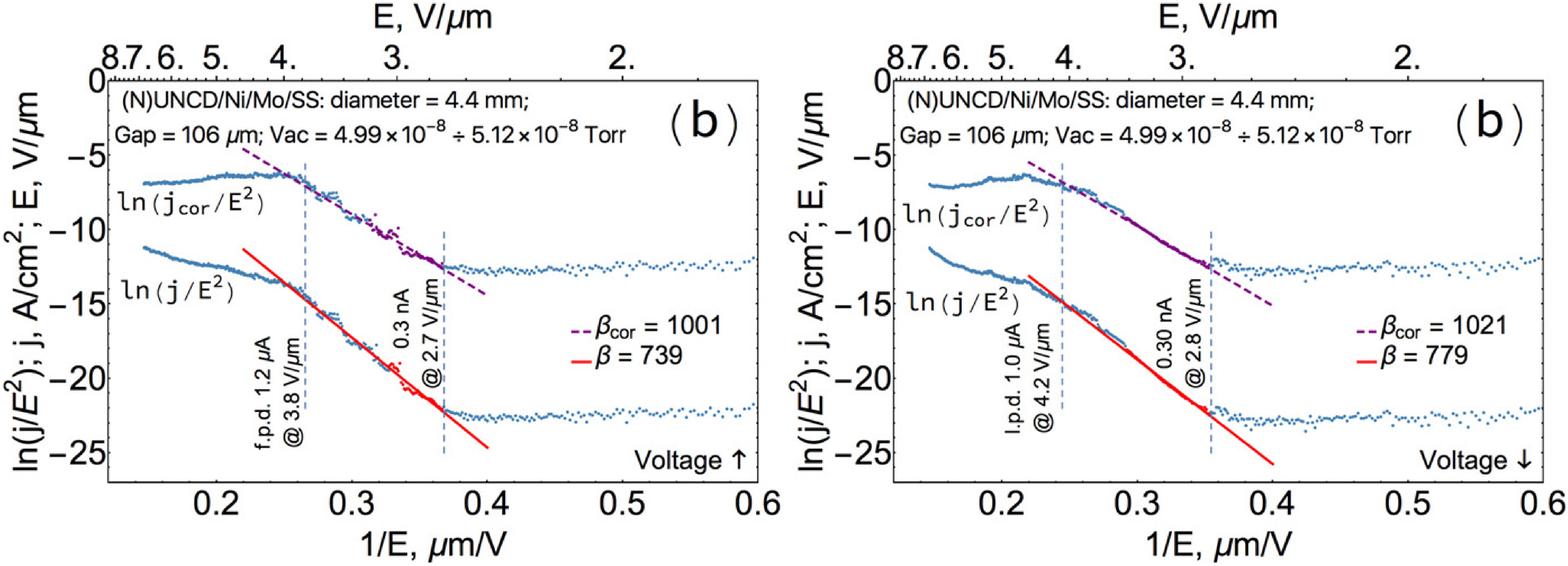}
\includegraphics[width=0.9 \textwidth]{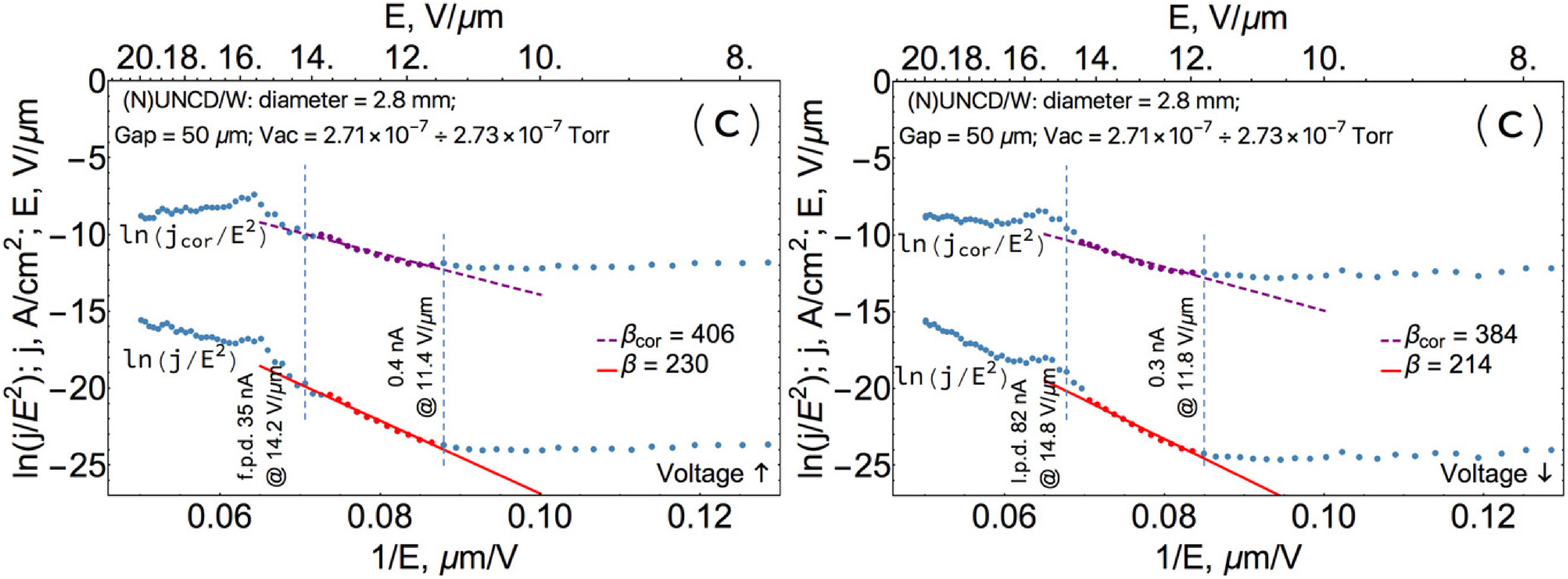}
\caption{Comparison between FN plots obtained by convention and by
normalizing with $S_{loc}(E)$ for (a) \textbf{NiMoSS1}, (b)
\textbf{NiMoSS2}, and (c) \textbf{W}. Left column combines ramp-up
FN plots. Right column are ramp-down FN plots. $j_{cor}$ are on
top and the convention $j$ are below.} \label{fig:8}
\end{figure*}

Most strikingly, it was observed that all $j_{cor}$-$E$ curves
saturated at $\sim$100 mA/cm$^2$, despite apparent significant
difference was seen from $I$-$E$ or $j$-$E$ curves. This leads us
to conclude that: (i) the saturation current level of $\sim$100
mA/cm$^2$ represents a basic intrinsic property of (N)UNCD films
while (ii) the $S_{loc}(E)$ value, i.e. the number of emitting
channels available on the surface, is an effect driven by the
substrate choice. The obtained results show that the deviation
from the FN law onsets (kink point) when the critical current
density is achieved regardless of the applied electric field. It
is demonstrated that the turn-on electric field threshold and
$\beta$-factor, representing the FN part of the $j$-$E$ curve, are
independent from the saturation current. This is summarized in
Fig.9. The sets \textbf{NiMoSS1} and \textbf{NiMoSS2} turned on at
2.5 V/$\mu$m while the set \textbf{W} required the field of 11-12
V/$\mu$m. This means macroscopic roughness (original roughness of
the W substrate, Fig.2a) and micro- and submicroscopic topography
(clustered spheres of the Ni/Mo/SS substrate, Fig.2b) are not
prerequisites for good FE properties. The substrate type is a key
factor under the default growth conditions in the CVD reactor.
This manifests in the Raman spectra (Fig.2d) as the decreased D/G
ratio, from standard 1.6 to 1.4-1.2, suggesting that emission
sites and their amount depend on the graphitic fraction in the
film. Similar effect was observed by other groups, see for
instance Ref. \cite{21} where the increased of amount of the
graphitic $sp^2$ phase led to improved field emission properties
of (N)UNCD films as was evaluated by Raman and NEXAFS
spectroscopy. The combined locally resolved field emission and
Raman microscopy study must be conducted to validate this
speculation.

The observed saturation effect places (N)UNCD together with the
rest of conventional semiconductors, in which identical effects
were reported as early as 1960's \cite{22,23}. It has been
consistently measured until this very day that, in contrast to
metals, the FN law breaks down. When plotted in FN,
$ln(j/E^2)$-$1/E$, coordinates, electric characteristics of
semiconductors deviate and rapidly part down from the straight
line. There are also a vast number of reports on the saturation
effect in carbon-based materials \cite{24,25,26,27}. The nature of
the saturation plateau was speculated to be due to electron
tunneling through multiple barriers in diamond films \cite{26}, or
most frequently due to the space charge effect \cite{27,28}.

At the same time, in the original work on the space charge effect
in field emission \cite{29}, current densities as high as 10$^7$
A/cm$^2$ are required to start screening the electric field.
Unless extreme localization of emitting centers is assumed, no
such current densities are observed in carbon-derived materials.
This let us to speculate that the saturation phenomenon in
amorphous carbon, polycrystalline diamond and CNT has the same
fundamental origin apart from the space charge effect. Further
experimental and theoretical work to explain the plateau and its
value $\sim$100 mA/cm$^2$, and investigation of this phenomenon
more thoroughly is underway. The developed methodology of
determining the actual emission area $S_{loc}(E)$ makes it
possible to study other carbonic field emitters and therefore,
e.g., to confirm or reject the hypothesis on the space-charge
effect in CNT fibers \cite{27}.

\begin{figure}[h]
\includegraphics[width=7.5cm]{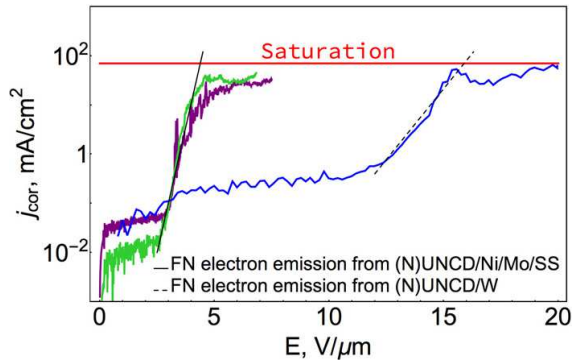}
\caption{Comparison plot of the sets \textbf{NiMoSS1} (purple
curve) \textbf{NiMoSS2} (green curve) and \textbf{W} (blue curve).
Black solid and dashed lines are to crudely depict FN part of the
$j$-$E$ curves.} \label{fig:9}
\end{figure}

There are two other consequences of the imaging experiments that
we have conducted. First, the electric characteristics plotted in
FN coordinates in Fig.8 suggest a certain increase of the
$\beta$-factors for the $j_{cor}$-$E$ curves family because the
pre-saturation region was modified after correcting with
$S_{loc}(E)$. Second, the hysteresis of $j$-$E$ characteristics in
Fig.5, which is commonly reported by many other groups (see e.g.
Ref.\cite{27}), is not caused by the hysteresis of the emission
area within accuracy of our experiment and image processing as we
do not observe significant hysteresis of the $S_{loc}(E)$, when
the voltage is ramped up and down as evidenced in Fig.6.

\section{IV. Summary}

An image processing concept and methodology were developed and
implemented, in order to extract the actual, locally resolved,
effective emission area of (N)UNCD films from emission pattern
micrographs acquired using a field emission projection microscope,
concurrently with current-voltage characteristics. It was shown
that the effective emission area depends non-linearly on the
applied electric field and repeatedly increases/decreases as the
field is ramped up/down.

By normalizing the measured current by the dynamic emission area
rather than by the constant entire cathode area and analyzing the
resulting $j$-$E$ curves, a few important conclusions were made:
(i) semi-metallic (N)UNCD saturates similarly as semiconductors;
(ii) in topographically uniform (N)UNCD thin films, field emission
is not uniform; (iii) field emission current density is limited at
$\sim$100 mA/cm$^2$; this is specific to (N)UNCD, regardless of
substrate; (iv) roughness and topography are not prerequisites for
good FE properties.

Additionally, the proposed concept of microscopy and image
processing has potential as a novel express technological
procedure to accurately quantify effects in as-synthesized
emitters and upon their surface termination/functionalization.

\

\noindent \textbf{Acknowledgments.} The authors would like to
thank Dr Igor Volkov (GWU) for his help with implementation of the
clustering algorithm and Dr. George Younes (GWU) for valuable
discussions.

Euclid TechLabs was supported by the Office of Nuclear Physics of
DOE through a Small Business Innovative Research grant \#DE-SC
0013145.

Samples synthesis, SEM and Raman measurements were conducted in
the Center for Nanoscale Materials at Argonne National Laboratory.
Use of the Center for Nanoscale Materials, an Office of Science
user facility, was supported by the U.S. Department of Energy,
Office of Science, Office of Basic Energy Sciences, under Contract
No. DE-AC02-06CH11357.

S.S. Baturin was supported by NSF grants \#PHY-1535639 and
\#PHY-1549132 during the time of preparing the manuscript for
publication.

\cleardoublepage

\noindent \textbf{Appendix A.} The automated image processing
algorithm used to process all datasets of the field emission
pattern micrographs.

\

The routine consists of the data preprocessing, local maxima
search, clustering, and segmentation. The implementation of each
of the steps is described as follows.

\subsection{A1. Data Preprocessing}
The first step towards the image processing is data preparation.
The input data is an actual RGB micrograph as shown in Fig.A1.
Built-in Mathematica functions, namely \textit{ColorConvert} and
\textit{ImageData}, are used to convert RGB images into grayscale,
represented in shades of gray between 0 (black) and 1 (white), and
to retrieve the obtained image data in the form of a 2D array of
pixel values (intensities), respectively. The intensity is
represented by $I_{i,j}$ with $i = 1, 2, \dotsc , i_{max}$, and $j
= 1, 2, \dotsc , j_{max}$, where $i_{max}$ and $j_{max}$ stand for
the image dimensions measured in pixels as shown in Fig.A2
(typically, our images are $442\times442$ pixels in size). As we
employ Poisson distribution for image processing, we can further
convert intensities into a 2D array of integer numbers $0, 1, 2,
\dotsc I_{max}$. To achieve this, each value in the dataset is
subtracted by the minimal intensity $I_{min}$ determined as
$Min(I_{i,j} | i = 1,\dotsc, i_{max}; j = 1,\dotsc, j_{max})$ and
divided by the minimal intensity difference in the array $\delta
I_{min}$ determined as $Min(\mathrm{Abs}(I_{m,n} - I_{l,k})|m, l =
1, \dotsc, i_{max}; n, k = 1, \dotsc , j_{max})$.

\renewcommand{\theequation}{A1}

\begin{align}
I_{i,j}=\frac{I_{i,j}-I_{min}}{\delta I_{min}}.
\end{align}

\setcounter{figure}{0}
\begin{figure}[h]
\renewcommand{\figurename}{FIG. A}
\includegraphics[width=5cm]{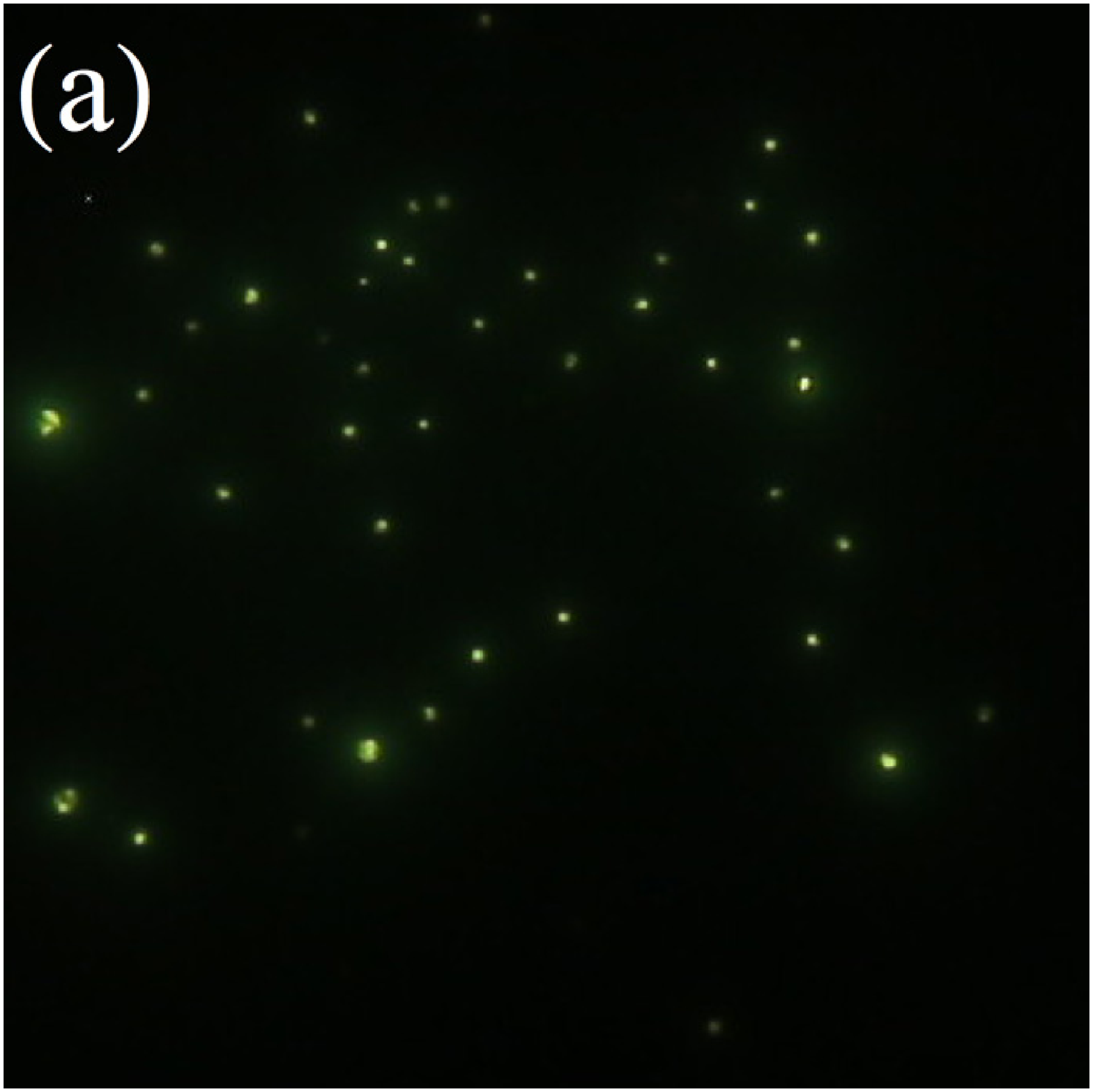}

\

\includegraphics[width=7cm]{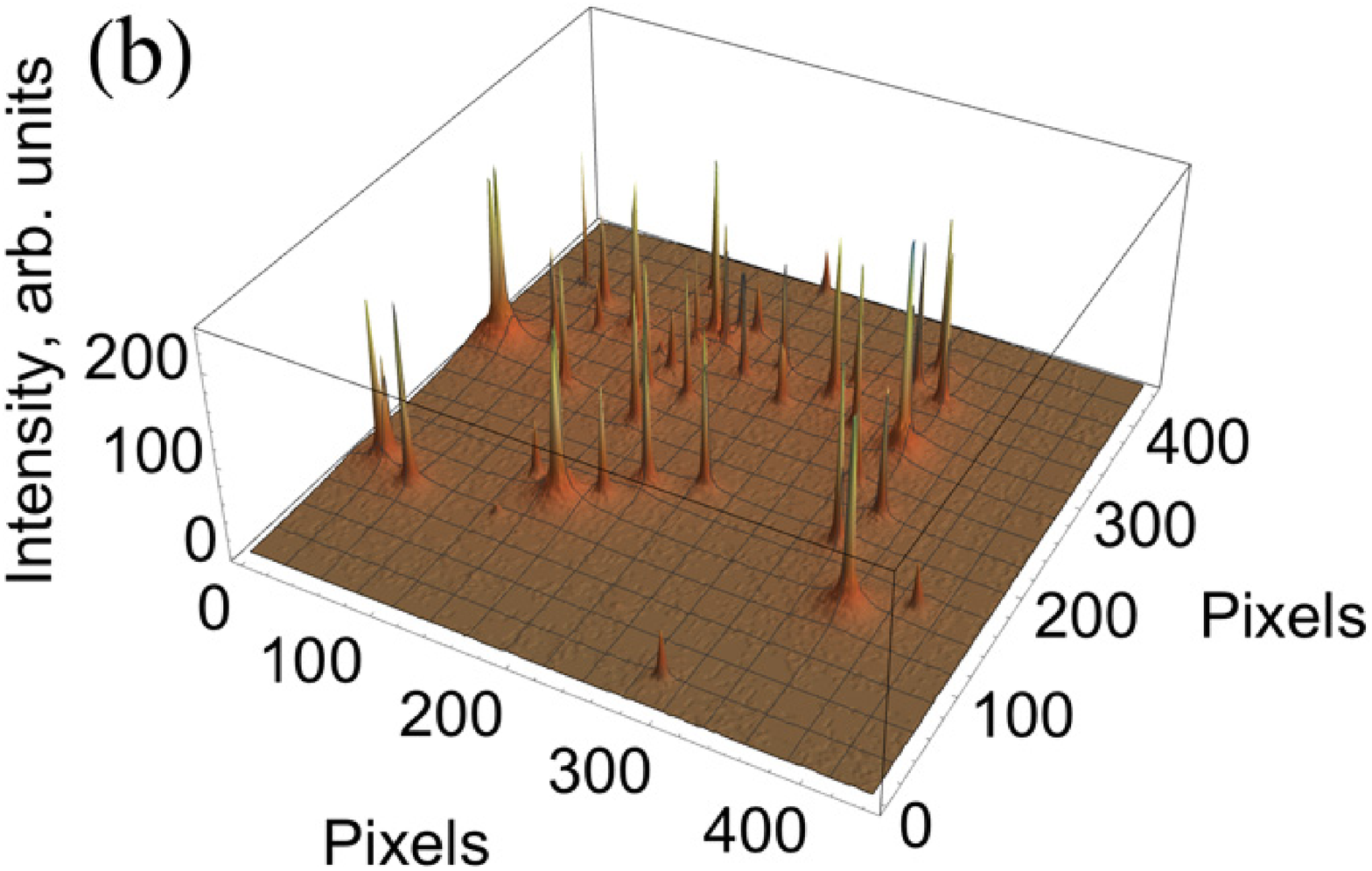}
\caption{(a) An example of a typical input image, field emission
pattern taken at the applied field strength of 5 V/$\mu$m during
$I-V$ measurements. (b) 3D plot of the resulting data array.}
\label{fig:A1}
\end{figure}

\begin{figure}[h]
\renewcommand{\figurename}{FIG. A}
\includegraphics[width=5cm]{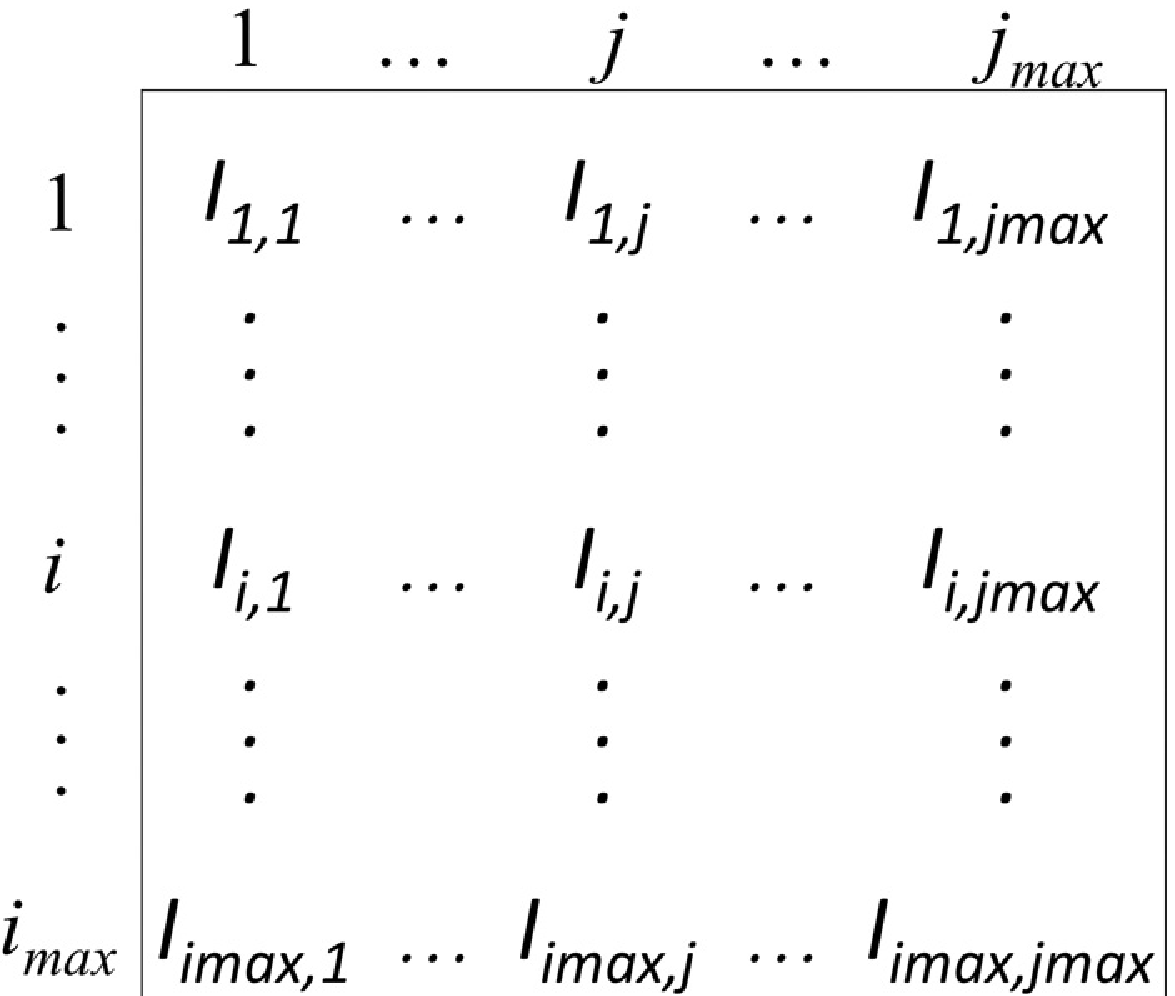}
\caption{Convention for labeling the indexes in the dataset of
pixel values $I_{i,j} = 0, 1, 2, \dots , I_{max}$.} \label{fig:A2}
\end{figure}

\subsection{A2. Local Maxima Search}
Here we use the idea that the local maximum is defined as a pixel
with an intensity larger than that of any other neighbor and
located at a relatively large distance from pixels with higher
intensities \cite{20}. The local maxima search routine involves
calculating the distance to the nearest pixel with higher
intensity. The distance is assumed to be the Euclidian distance
$r$ between two pixels $I_{m,n}$ and $I_{l,k}$ with coordinates
$(m, n)$ and $(l, k)$, respectively
\renewcommand{\theequation}{A2}
\begin{align}
r=\sqrt{(m-l)^2+(n-k)^2}
\end{align}

\noindent with $m$ and $l$ running over 1 to $i_{max}$, and $n$
and $k$ running over 1 to $j_{max}$. This procedure is
accomplished for each pixel $I_{i,j}$ by examining the intensities
of all neighbor pixels, which lie inside the circle of radius
$r_{cr}$ centered at the particular pixel. In our case, for a
$442\times442$ pixel array, we use $r_{cr} = 10$. To illustrate
the general approach, let us consider a small part of a dataset.
Fig.A3 illustrates the closest brighter pixel search in a
$15\times15$ pixel array. For most pixels, the nearest brighter
pixels will lie at a distance $r = 1$. For example, Fig.A3a shows
the closest brighter pixel search for the pixel $I_{1,1}$, where
we highlight in green the search area. It can be seen that the
closest brighter pixel $I_{2,1} = 6$ is located at $r = 1$,
whereas pixel $I_{6,4} = 136$ does not have any brighter pixels in
the highlighted neighborhood (Fig.A3b). We define such pixels as
\emph{the brightest in the neighborhood}. Next, for each of the
brightest in the neighborhood pixels, we calculate the distance to
the closest brighter pixel in the list of the brightest pixels.
For the pixel with highest intensity (global maximum), we take the
distance to the closest pixel in the list of the brightest pixels.
At the end of this routine, each pixel of a 2D array is associated
with its own intensity $I_{i,j}$ and distance to the nearest
brightest pixel $r_{i,j}$.

\begin{figure}[h]
\renewcommand{\figurename}{FIG. A}
\includegraphics[width=7cm]{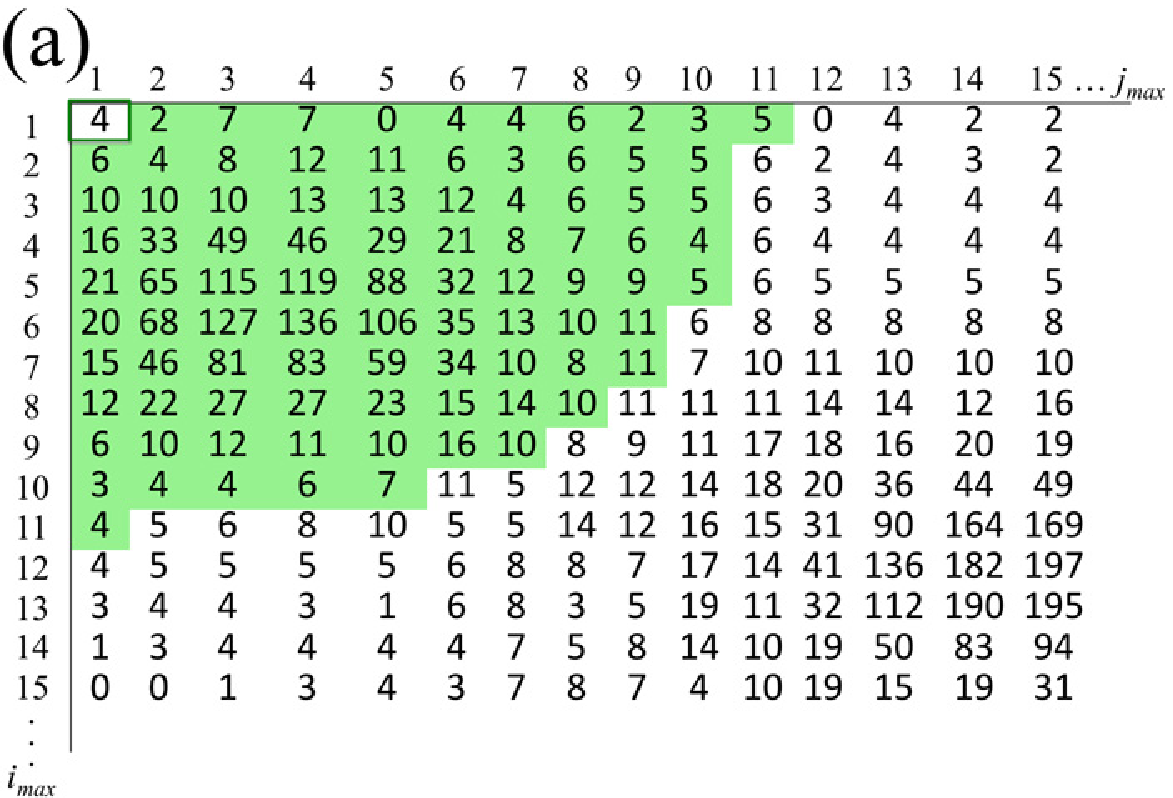}

\

\includegraphics[width=7cm]{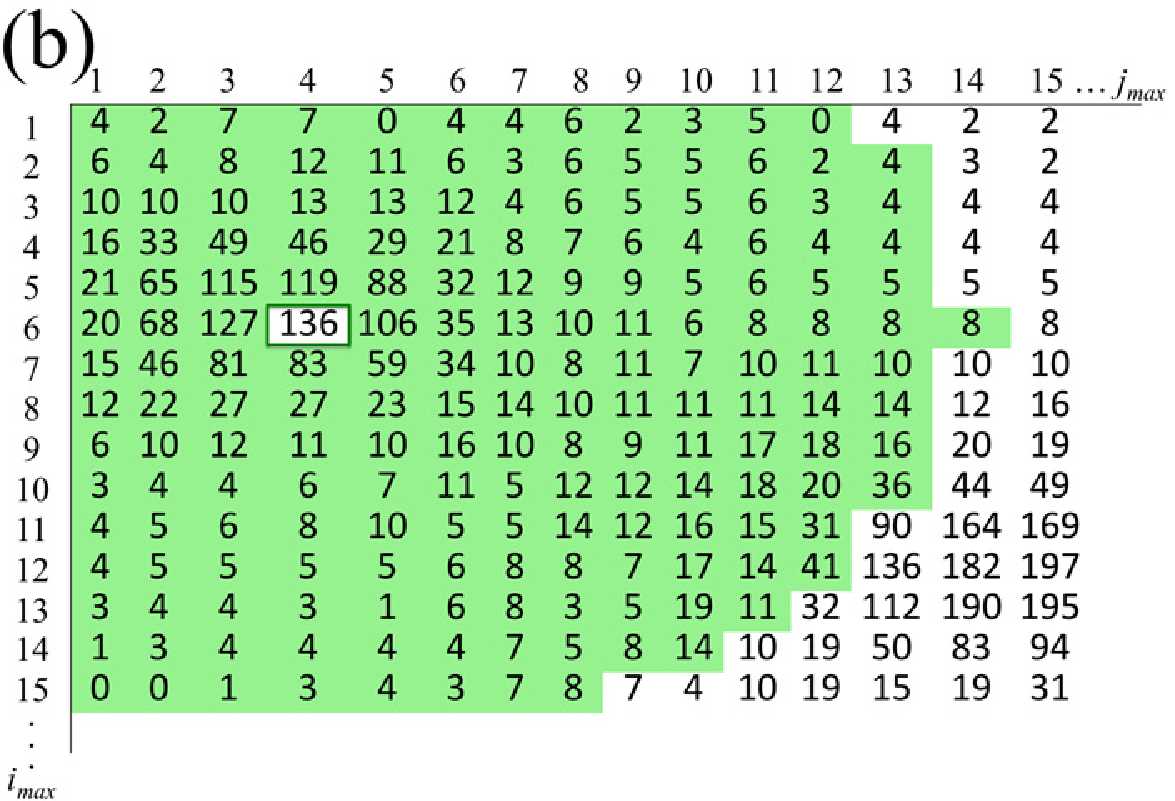}

\caption{Illustration of the closest brighter pixel search
algorithm. (a) Most of the pixels have their brighter neighbors at
relatively small distances. Pixel $I_{1,1} = 4$ has its closest
brighter neighbor $I_{2,1} = 6$ located at $r = 1$. (b) Pixel
$I_{6,4} = 136$ is the brightest in the neighborhood. The closest
brighter pixel $I_{12,15} = 197$ is located at $r = 12.53$, which
is out of $r_{cr}$. The pixel $I_{12,15}$ is also defined as the
brightest in its neighborhood.} \label{fig:A3}
\end{figure}

It is convenient to plot the distance $r_{i,j}$ as a function of
intensity $I_{i,j}$ for each pixel. We will further call this
representation the decision plot as defined by Rodriguez and Laio
\cite{20}. Fig.A4 shows the decision plot for the image in
Fig.A1a. It can be seen that only some of the pixels possess a
combination of high intensity and a distance much larger than
$r_{cr}$. These are \textit{local maxima}.

In order to completely automate the local maxima search, we need
to define two parameters, the minimal distance $D_{min}$ and the
minimal intensity $I_{min}$. The pixel, which possesses these
values, will be classified as a local maximum. The choice of
$D_{min}$ primarily depends on the image resolution. Images with
higher resolution will require a larger value of this parameter.
It was found that for $442\times442$ pixel image, $D_{min}$ = 7 is
an appropriate choice, whereas the choice of $I_{min}$ requires an
estimation of the background level.

\begin{figure}[h]
\renewcommand{\figurename}{FIG. A}
\includegraphics[width=7cm]{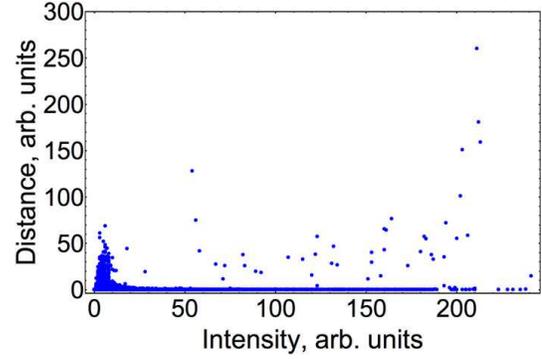}
\caption{An example of the decision plot for the image shown in
\ref{fig:A1}a.}
\label{fig:A4}
\end{figure}

Each pixel intensity $I_{i,j}$ is a sum of both the true signal,
the light emission from a discrete emission site, the background,
the emission from other emission sites, and the noise contributed
by the light detection system itself. One common method of
separating the signal from the background is based on an
assumption that the background follows a Poisson distribution
\cite{16,19}. The probability of observing a pixel with a random
intensity $I_{i.j}$, when the mean intensity is $\lambda$, can be
written as
\renewcommand{\theequation}{A3}
\begin{align}
P_{i,j}=\frac{\lambda^{I_{i,j}}e^{-\lambda}}{I_{i,j}!}.
\end{align}
Pixels that satisfy the condition
\renewcommand{\theequation}{A4}
\begin{align}
\label{eq:cond}
P_{i,j}\leq 0.01/N,
\end{align}
where $N$ is the total number of pixels, are assumed to be the
meaningful signal and are subject to further examination
\cite{16}. Since there are several emission sites on the picture,
each with a different brightness, the background level can be
overestimated. Therefore, pixels that satisfy the above condition
are removed from the dataset and the procedure is repeated until
there are no additional pixels of interest. The background level
is then determined by the largest element of a final data list.
The histogram in Fig.A5 illustrates an automatically determined
background level. Pixels with $I_{i,j} \geq 23$ and $r_{i,j} \geq
7$ are local maxima and are marked in red on the decision plot
(Fig.A6).

\begin{figure}[h]
\renewcommand{\figurename}{FIG. A}
\includegraphics[width=7cm]{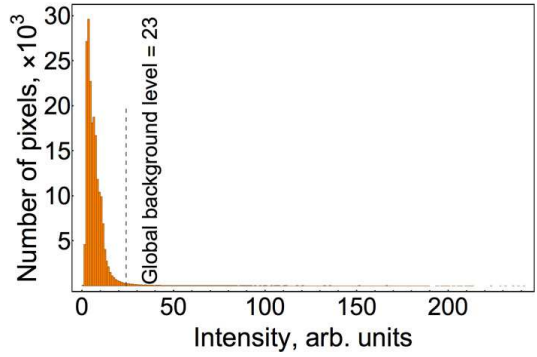}
\caption{Histogram of data for the image in Fig.A1a. The algorithm
picks pixels until the condition (A4) holds true.}
\label{fig:A5}
\end{figure}

\begin{figure}[h]
\renewcommand{\figurename}{FIG. A}
\includegraphics[width=7cm]{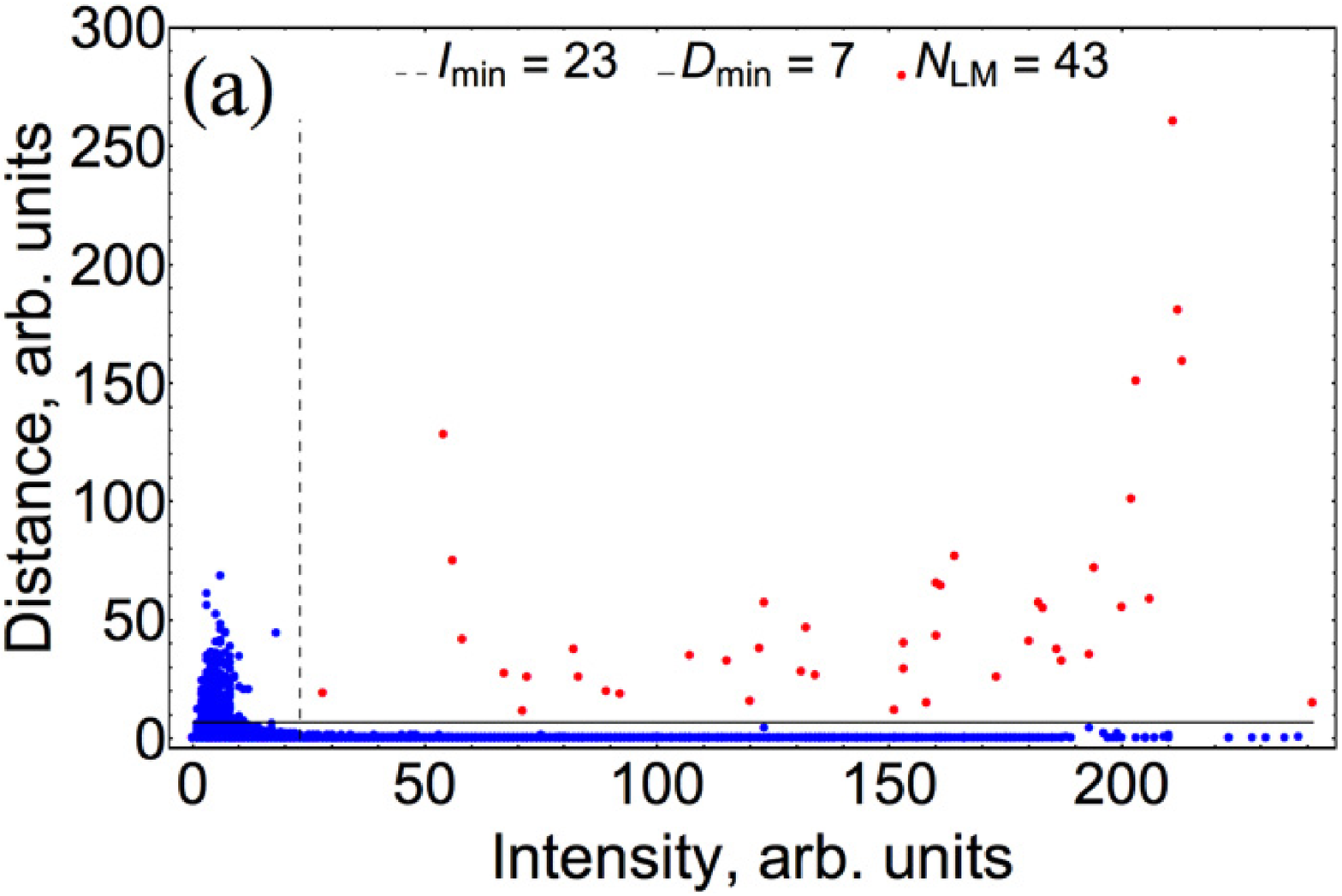}

\

\includegraphics[width=7cm]{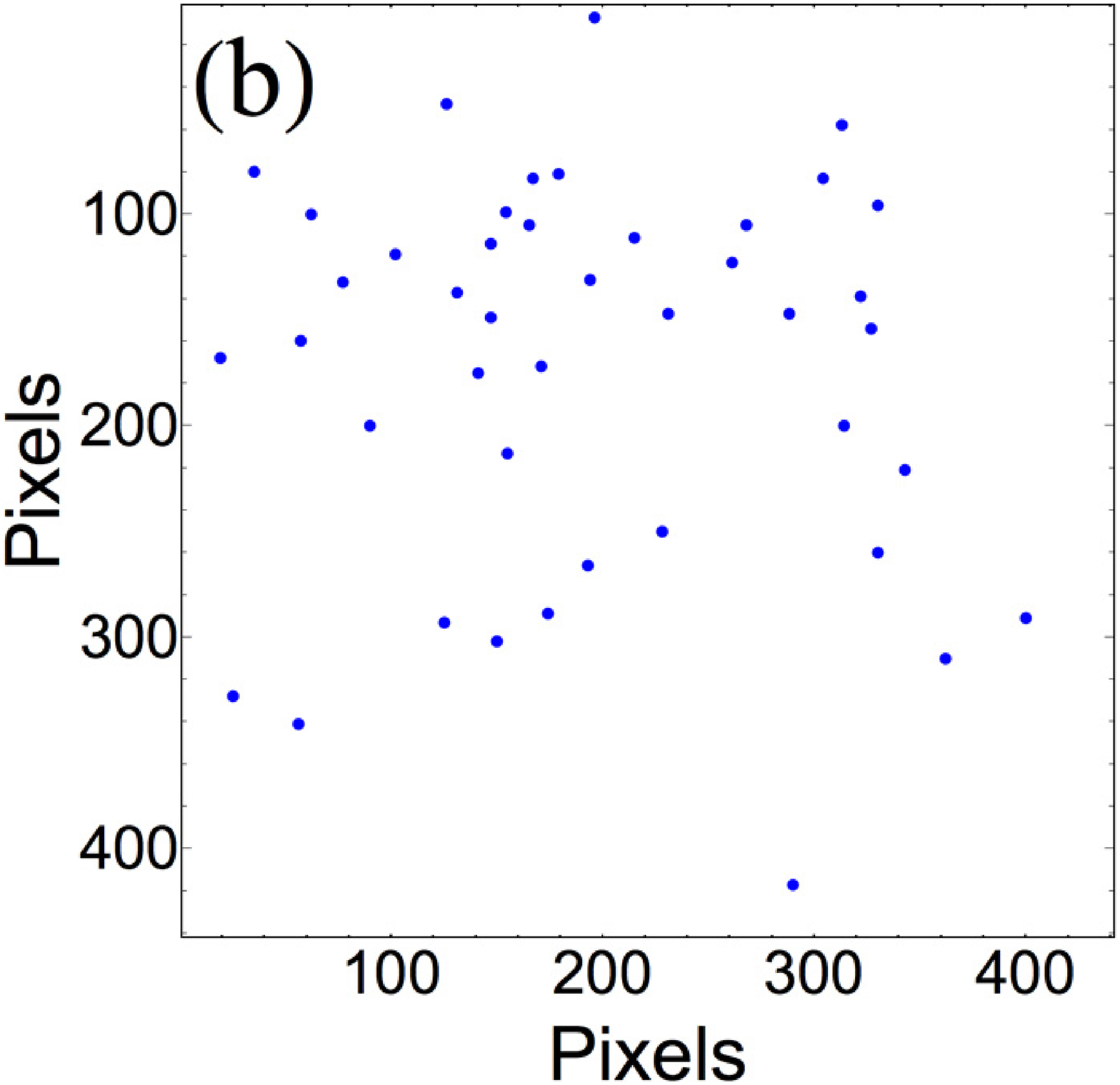}

\caption{(a) Decision plot for the image presented in Fig.A1a with
automatically calculated number of local maxima $N_{LM}$. Red dots
correspond to Poisson distribution of a final dataset (background
noise). (b) The pixel map of the found local maxima.}
\label{fig:A6}
\end{figure}

\subsection{A3. Clustering}
The methodology described above determines the number of local
maxima, i.e. the number of seeding centers around which clusters
will be built. The clustering procedure is performed by linking
the pixels around the local maxima together. For each local
maximum, the algorithm seeks the pixels for which this particular
local maximum is a brighter pixel and attaches these pixels to the
local maximum. Attached pixels, in turn, have pixels for which
they are brighter neighbors (Fig.A7). The procedure is repeated
for each local maximum until no more corresponding pixels are
found. At this point, clusters are complete. An example of a
$442\times442$ pixel independent cluster family is shown in
Fig.A8. Independent clusters around local maxima are color-coded.

\begin{figure}[h]
\renewcommand{\figurename}{FIG. A}
\includegraphics[width=7cm]{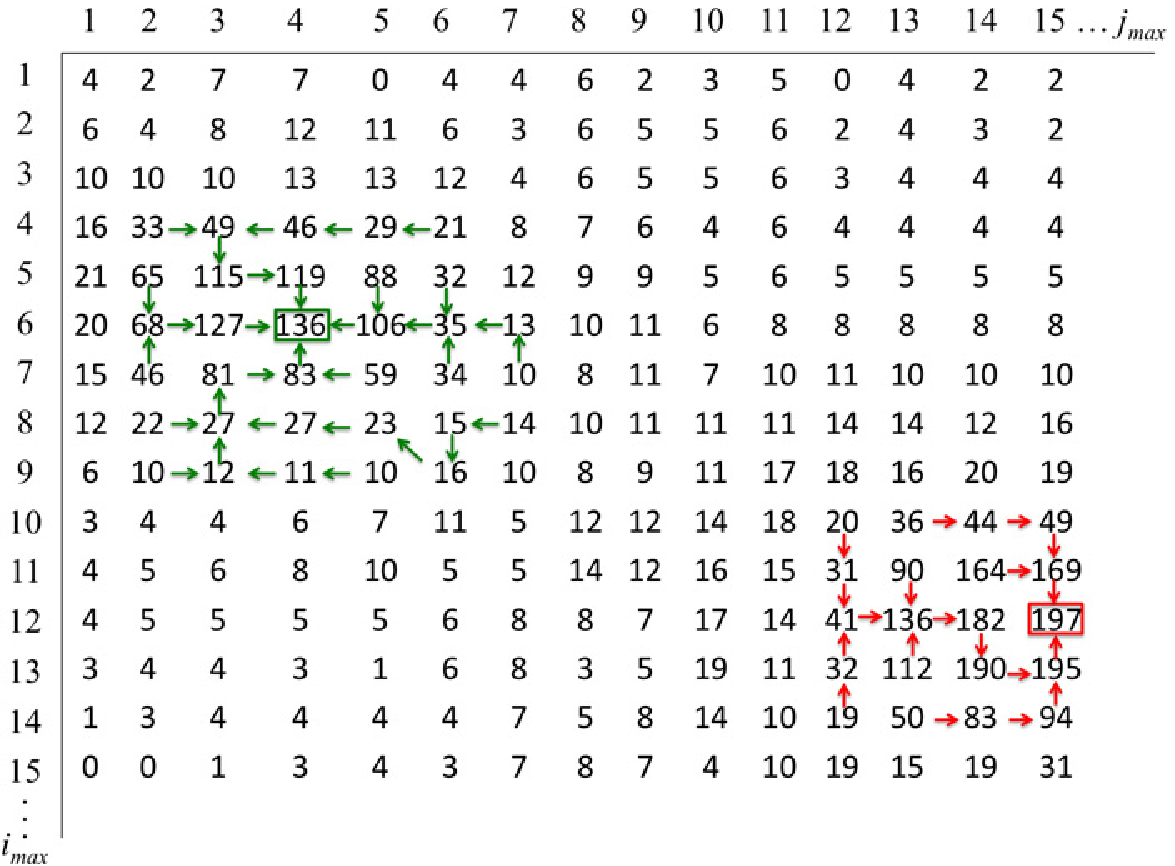}
\caption{Illustration of a clustering algorithm. Different colors represent a formation of two different clusters around two local maxima.}
\label{fig:A7}
\end{figure}

\begin{figure}[h]
\renewcommand{\figurename}{FIG. A}
\hspace*{-1.5cm} \includegraphics[width=7.5cm]{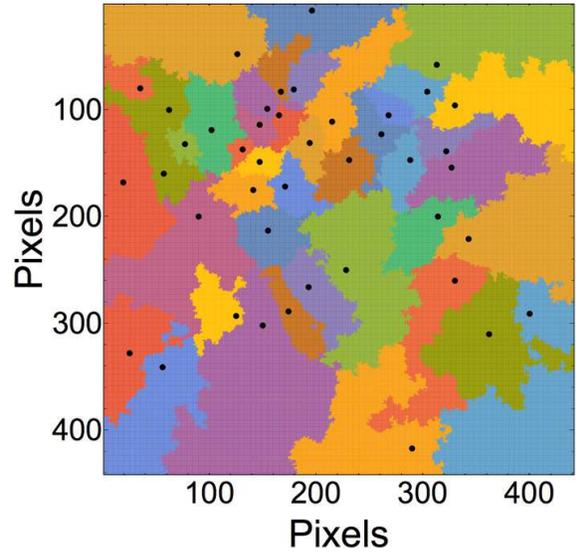}
\caption{Final result of a clustering procedure.} \label{fig:A8}
\end{figure}

\subsection{A4. Segmentation}
After clusters are formed around each local maximum, we perform a
segmentation procedure to decide how many pixels belong to each
emission site and therefore to evaluate the total emission area at
a given electric field. In order to estimate a local background,
we assume that each emission site has a Gaussian shape and use a
2D Gaussian function to fit each peak
\renewcommand{\theequation}{A5}
\begin{align}
f(x,y)=Ae^{-\frac{(x-x_0)^2}{2\sigma_x^2}-\frac{(y-y_0)^2}{2\sigma_y^2}}+B_{loc}.
\end{align}
Here coefficients $A$ and $B_{loc}$ are the amplitude and local
background, respectively, $x_0$ and $y_0$ define the pixel
position of a local maximum, and $\sigma_x$ and $\sigma_y$ are the
standard deviations in perpendicular directions (Fig.A9).

\begin{figure}[h]
\renewcommand{\figurename}{FIG. A}
\includegraphics[width=7cm]{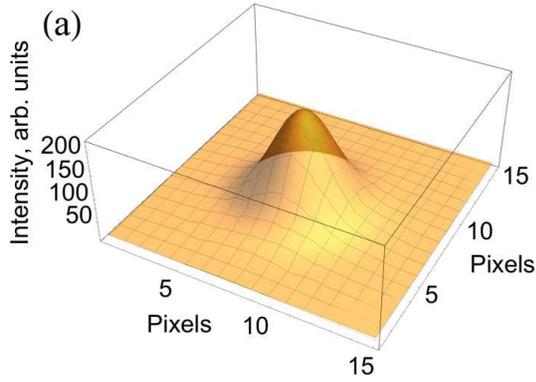}

\

\includegraphics[width=7cm]{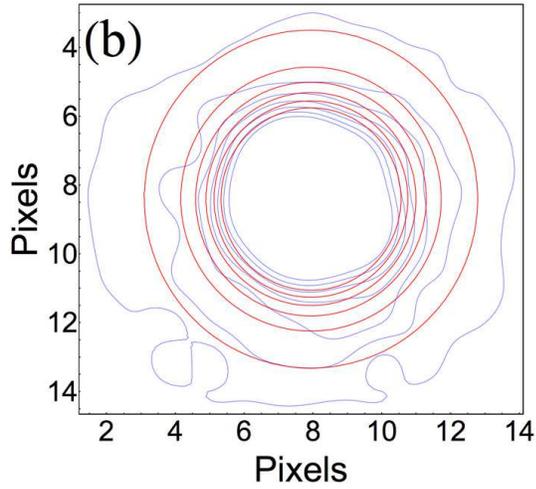}

\caption{(a) An emission peak with a fitting Gaussian curve and
(b) a corresponding contour plot.} \label{fig:A9}
\end{figure}

Further, we select only the pixels with intensity within one
standard deviation away from the mean, which means that only the
pixels that exceed the threshold
\renewcommand{\theequation}{A6}
\begin{align}
I_{th}=0.61I_{max}+0.4B_{loc}
\end{align}
contribute to the emission area. Here $I_{max}$ is the local
maximum intensity and $B_{loc}$ is the local background (see
Fig.A10).

\begin{figure}[h]
\renewcommand{\figurename}{FIG. A}
\includegraphics[width=7cm]{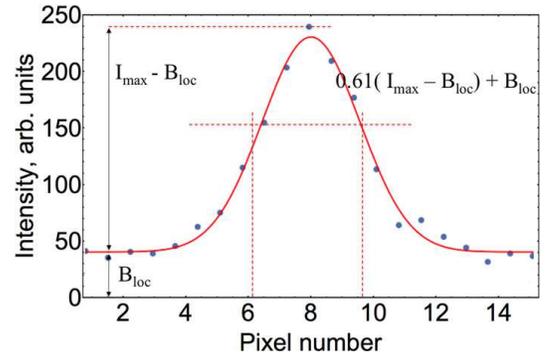}
\caption{Result of a clustering algorithm.}
\label{fig:A10}
\end{figure}

The pixel size in cm$^2$ is referenced to the full image size in
pixels and the known diameter of the (N)UNCD cathode, 0.44 or 0.28
cm. The product of the total number of pixels combined by the
clustering procedure and the pixel size in cm$^2$ provides the
total emission area per micrograph in the datasets shown in the
Appendices B to E.

\cleardoublepage

\begin{widetext}
\noindent\textbf{Appendix B.} The micrograph set for (N)UNCD/Mo/SS
measured at an inter-electrode gap (UNCD-YAG) of 130 $\mu$m and
pressure $1.65\times10^{-8}$ Torr. All 11 images were acquired in
the course of ramping the voltage up.
\begin{center}
\includegraphics[width=0.8 \textwidth]{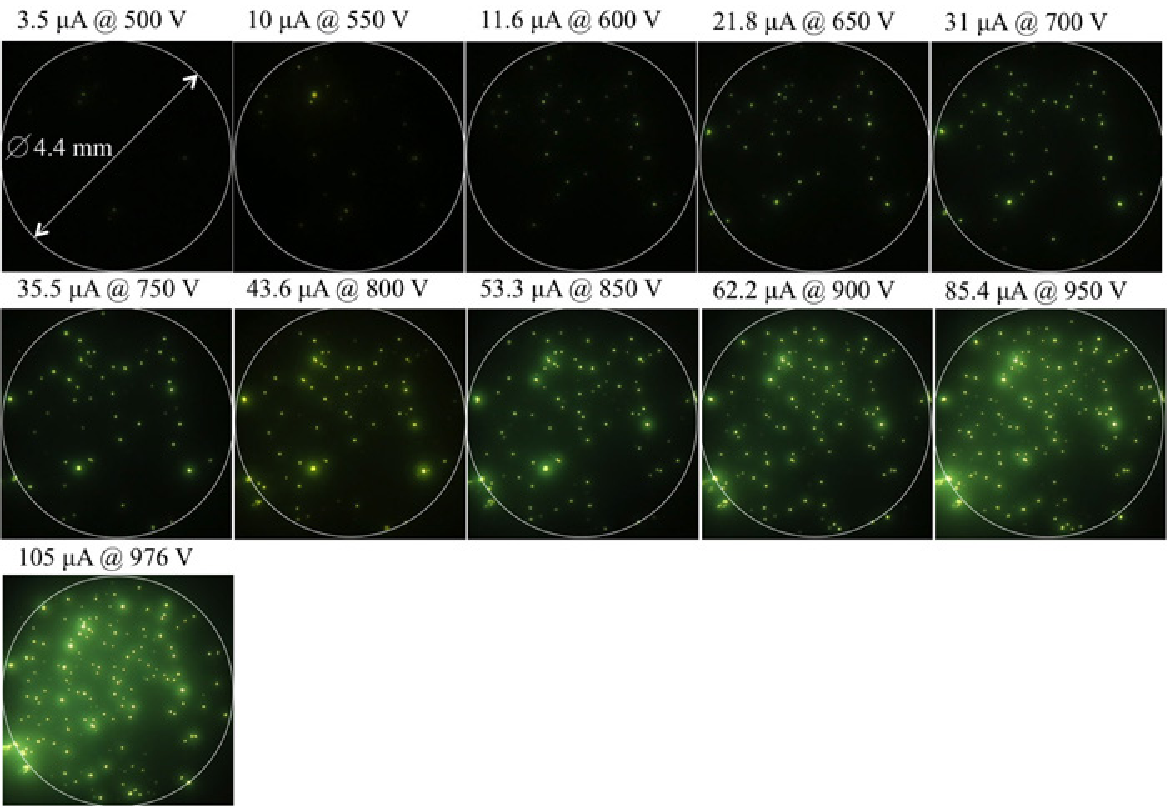}
\end{center}

\cleardoublepage

\noindent\textbf{Appendix C.} The micrograph set for
(N)UNCD/Ni/Mo/SS measured at an inter-electrode gap (UNCD-YAG) of
147 $\mu$m and pressure about $9\times10^{-8}$ Torr. The presented
51 images are those on which the image processing algorithm has
detected high intensity pixels above the background threshold. The
full set was recorded in the course of ramping the voltage up (0
to 1,100 V with a step of 20 V) and then down (1,100 to 0 V with a
step of 20 V).
\begin{center}
\includegraphics[width=0.8 \textwidth]{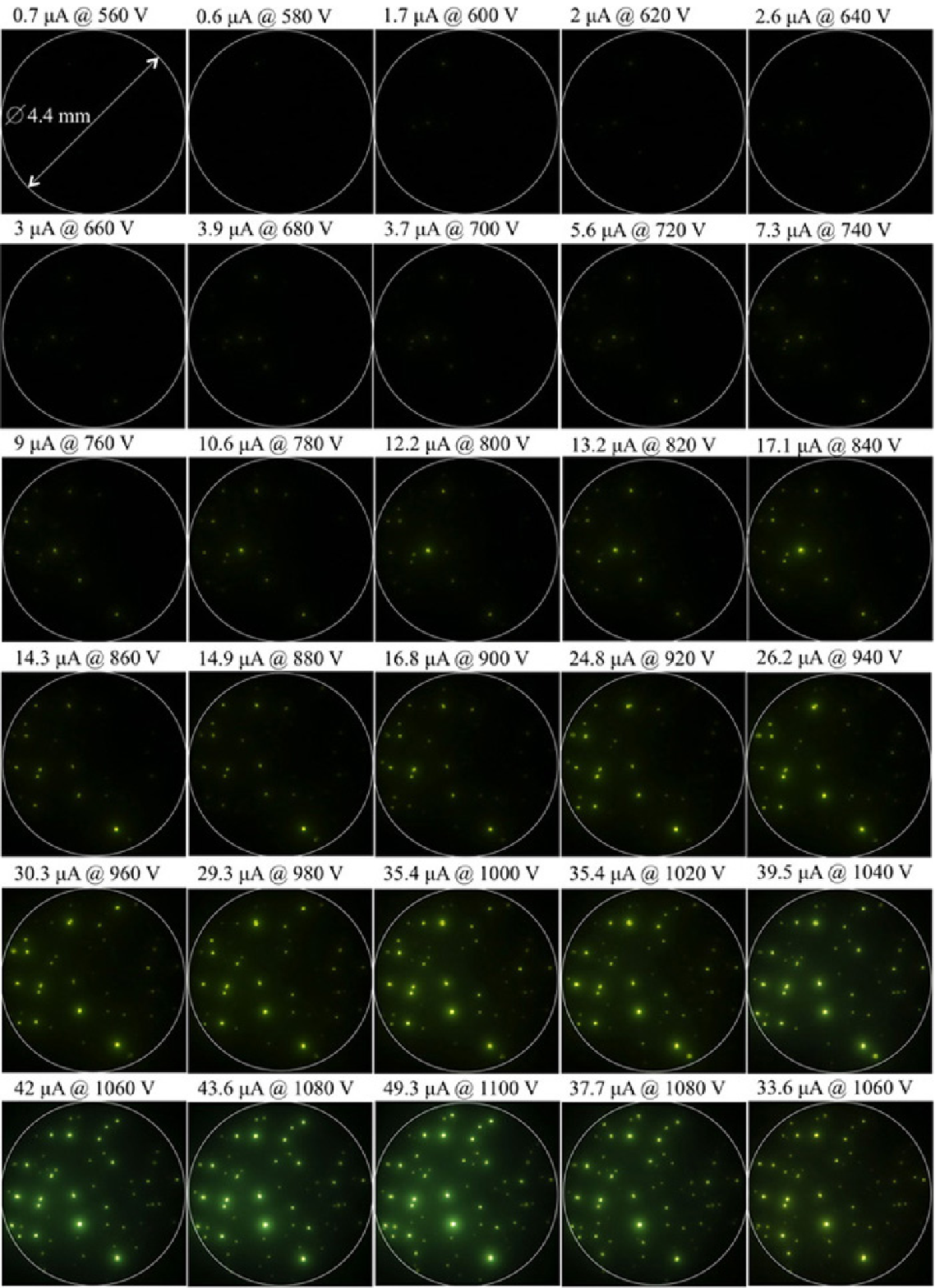}{\centering}
\end{center}

\begin{center}
\includegraphics[width=0.8 \textwidth]{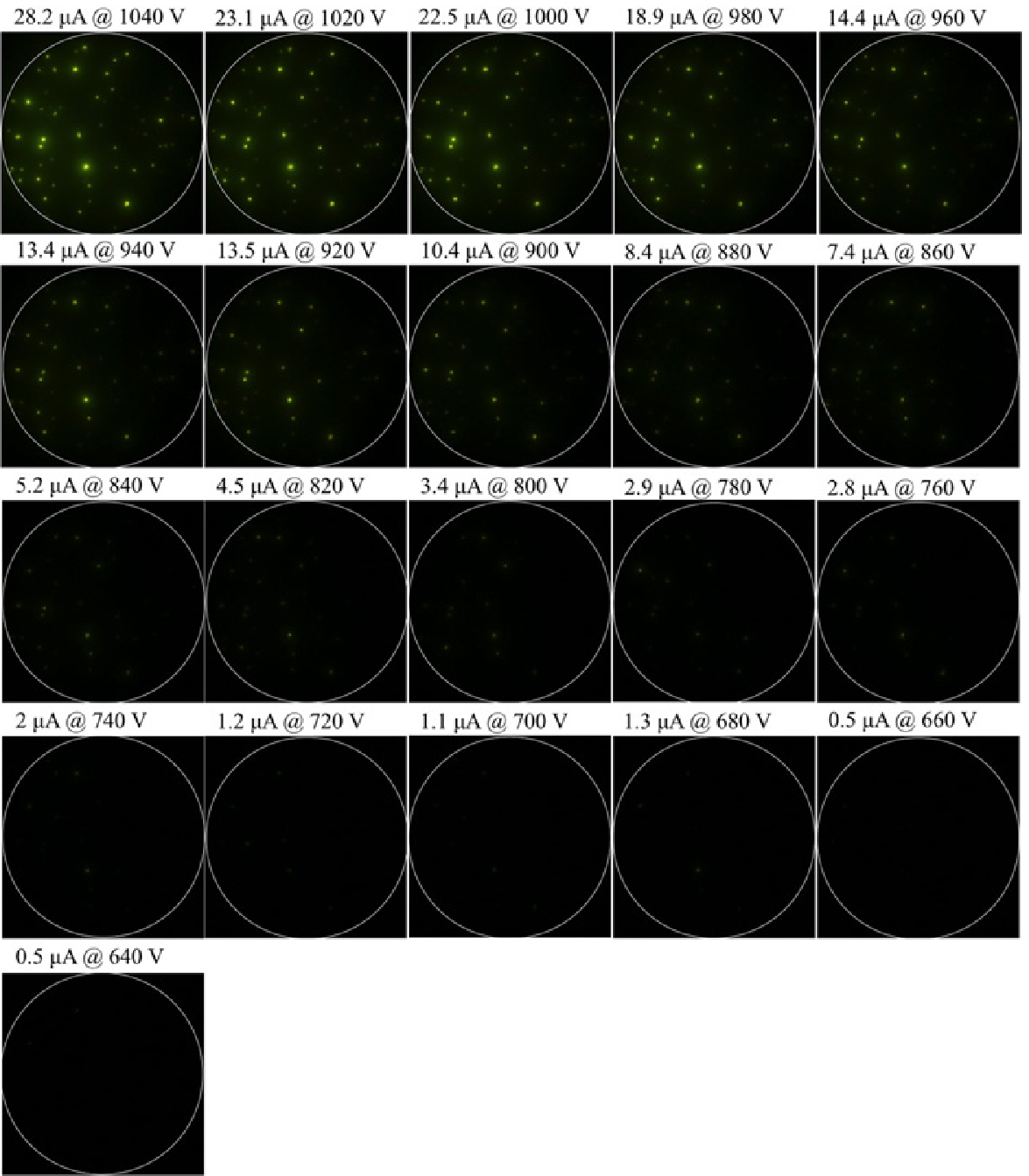}
\end{center}

\cleardoublepage

\noindent\textbf{Appendix D.} The micrograph set for
(N)UNCD/Ni/Mo/SS measured at an inter-electrode gap (UNCD-YAG) of
106 $\mu$m and pressure about $5\times10^{-8}$ Torr. The presented
30 images are those on which the image processing algorithm has
detected high intensity pixels above the background threshold. The
full set was recorded in the course of ramping the voltage up (0
to 720 V with a step of 20 V) and then down (720 to 0 V with a
step of 20 V).
\begin{center}
\includegraphics[width=0.8 \textwidth]{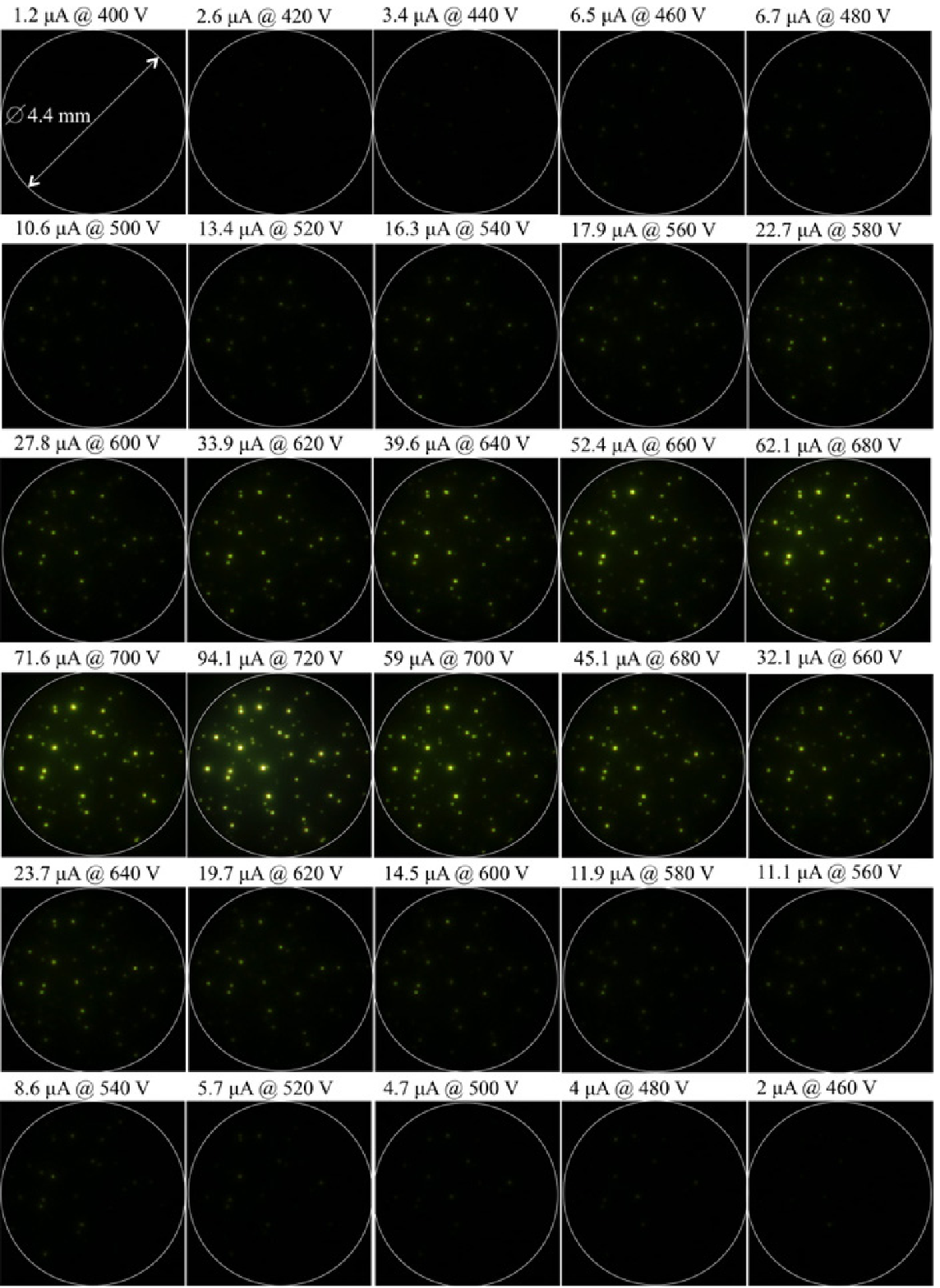}
\end{center}

\cleardoublepage

\noindent\textbf{Appendix E.} The micrograph set for (N)UNCD/W
measured at an inter-electrode gap (UNCD-YAG) of 50 $\mu$m and
pressure about $3\times10^{-7}$ Torr. The presented 56 images are
those on which the image processing algorithm has detected high
intensity pixels above the threshold. The full set was recorded in
the course of ramping the voltage up (0 to 1,000 V with a step of
10 V) and then down (1,000 to 0 V with a step of 10 V).

\begin{center}
\includegraphics[width=0.8 \textwidth]{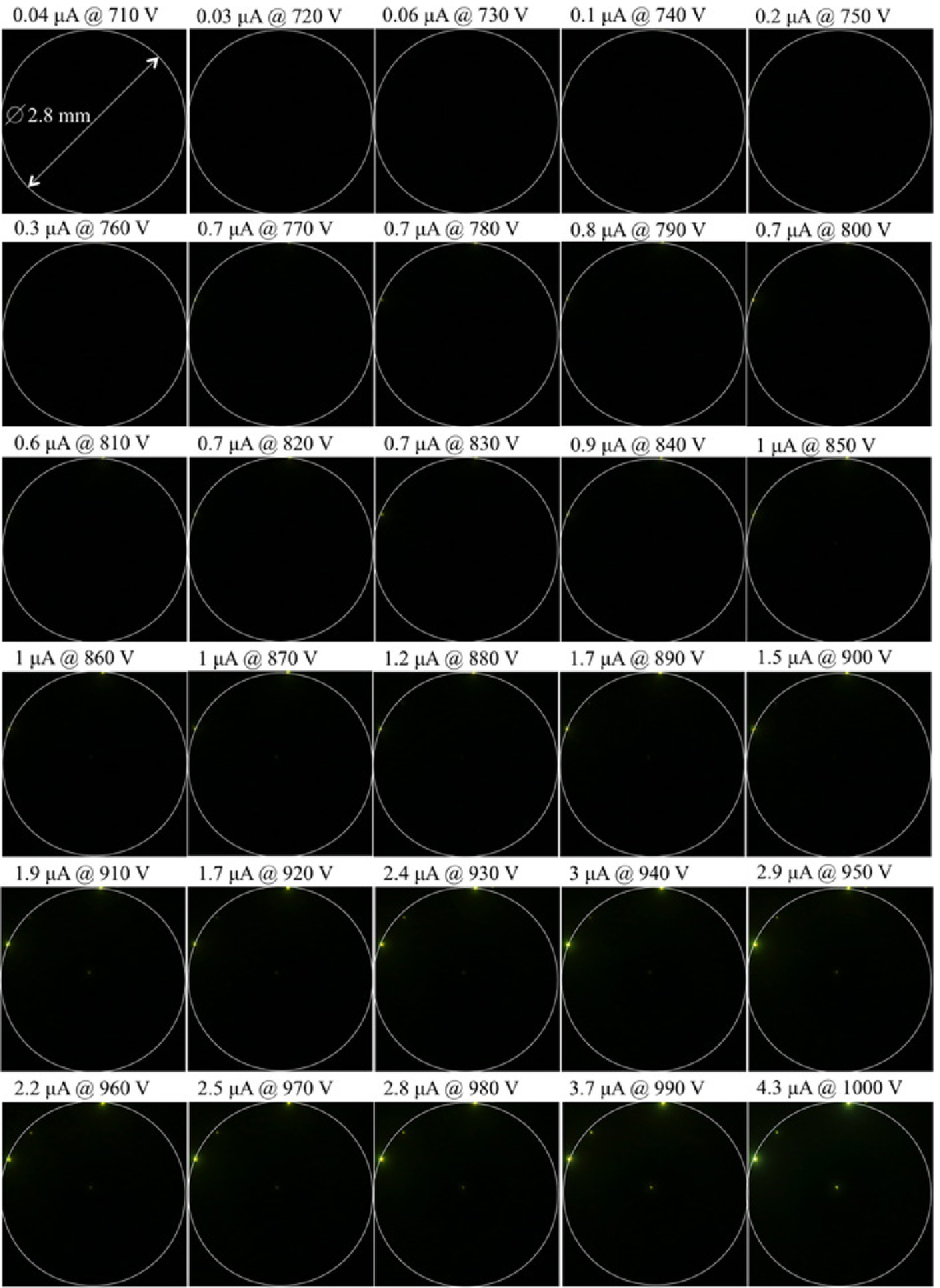}
\end{center}

\begin{center}
\includegraphics[width=0.8 \textwidth]{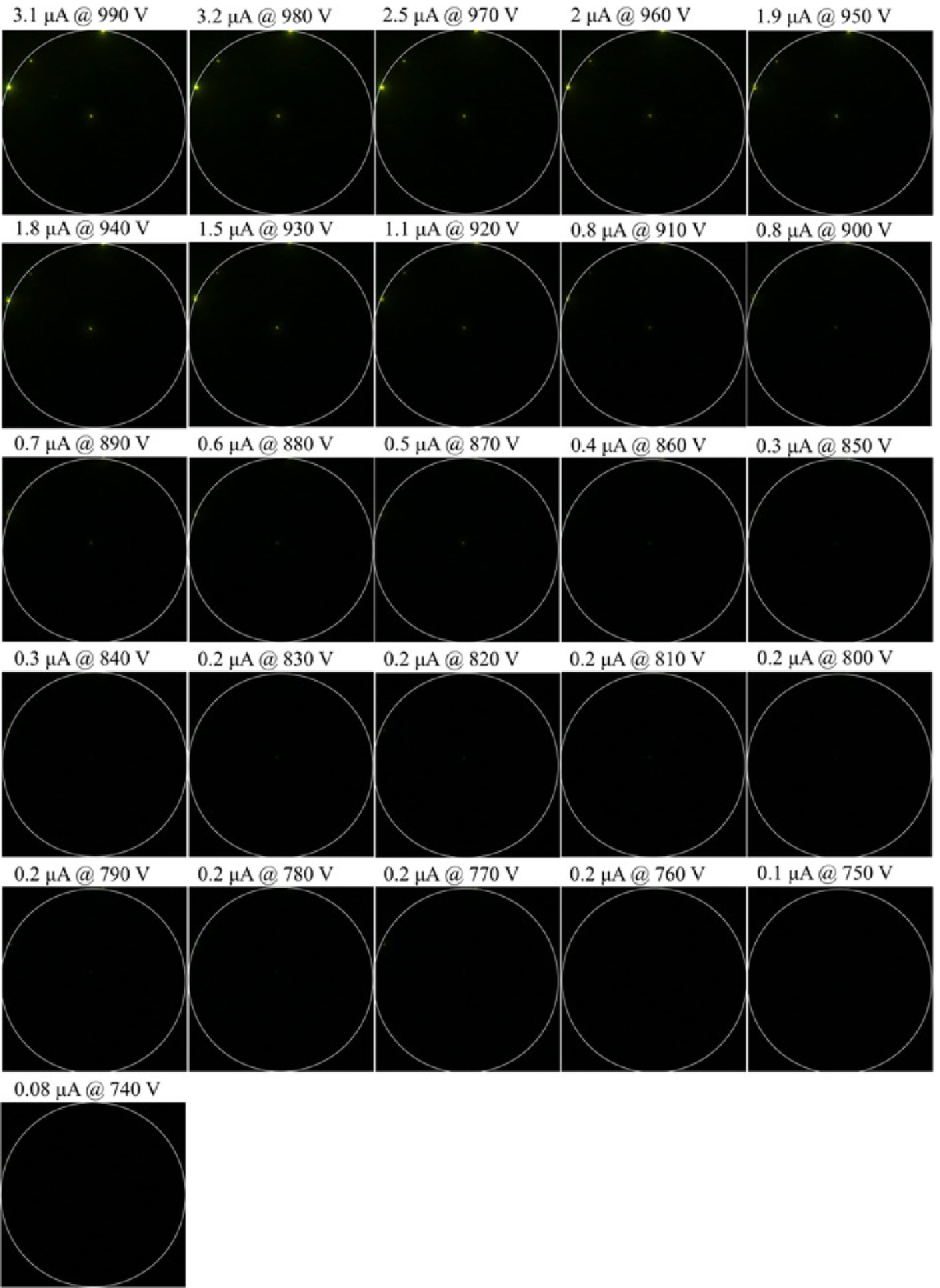}
\end{center}

\end{widetext}


\begin{thebibliography}{29}
\bibitem{1} R.H. Fowler and L. Nordheim, Proc. Royal Soc. A \textbf{119}, 173 (1928).

\bibitem{2} P. Serbun, A systematic investigation of carbon, metallic and semiconductor nanostructures for field-emission cathode applications,
Ph.D. thesis, University of Wuppertal (2014).

\bibitem{3} P. Serbun, B. Bornmann, A. Navitski, G. Müller, C. Prommesberger, C. Langer, F. Dams, and R. Schreiner, J. Vac. Sci. Technol. B \textbf{31}, 02B101 (2013).

\bibitem{4} R. Schreiner, C. Langer, C. Prommesberger, S. Mingels, P. Serbun, and G. Muller, IEEE Proc. \textbf{doi:}10.1109/IVNC.2013.6624721 (2013).

\bibitem{5} D. Lysenkov and G. Muller, International J. Nanotechnology \textbf{2}, 239 (2005).

\bibitem{6} A.G. Kolosko, E.O. Popov, S.V. Filippov, and P.A. Romanov, J. Vac. Sci. Technol. B \textbf{33}, 03C104 (2015).

\bibitem{7} S. A. Lyashenko, A. P. Volkov, R. R. Ismagilov, and A. N. Obraztsov, Tech. Phys. Lett. \textbf{35}, 249 (2009).

\bibitem{8} M.L. Terranova, S. Orlanducci, M. Rossi, and E. Tamburri, Nanoscale \textbf{7}, 5094 (2015).

\bibitem{9} A.V. Sumant, O. Auciello, R.W. Carpick, S. Srinivasan, and J.E. Butler, MRS Bull. \textbf{35}, 281 (2011).

\bibitem{10} S.S. Baturin and S.V. Baryshev, Rev. Sci. Instrum. \textbf{88}, 033701 (2017).

\bibitem{11} K.J. P\'{e}rez Quintero, S. Antipov, A.V. Sumant, C. Jing, and S.V. Baryshev, Appl. Phys. Lett. \textbf{105}, 123103 (2014).

\bibitem{12} S.V. Baryshev, S. Antipov, J. Shao, C. Jing, K.J. Pérez Quintero, J. Qiu, W. Liu, W. Gai, A.D. Kanareykin, and A.V. Sumant,
Appl. Phys. Lett. \textbf{105}, 203505 (2014).

\bibitem{13} J. Birrell, J.E. Gerbi, O. Auciello, J.M. Gibson, J. Johnson, and J.A. Carlisle, Diamond Relat. Mater. \textbf{14}, 86 (2005).

\bibitem{14} C.-C. Teng, S.-M. Song, C.-M. Sung, and C.-T. Lin, J. Nanomaterials \textbf{2009}, 621208 (2009).

\bibitem{15} F. Klauser, D. Steinmüller-Nethl, R. Kaindl, E. Bertel, and N. Memmel, Chemical Vapor Deposition \textbf{16}, 127
(2010).

\bibitem{16} F.P. Gavriil, V.M. Kaspi, and P.M. Woods, Astrophys. J. \textbf{607}, 959 (2004).

\bibitem{17} P.-K. Chan, S.-H. Cheng, and T.-C. Poon, J. Electronic Imaging \textbf{16}, 043003 (2007).

\bibitem{18} A. Khotanzad and A. Bouarfa, Pattern Recognition \textbf{23}, 961 (1990).

\bibitem{19} A.F. Zepka, J.M. Cordes, and I. Wasserman, Astrophys. J. \textbf{427}, 438 (1994).

\bibitem{20} A. Rodriguez and A. Laio, Science \textbf{344}, 1492 (2014).

\bibitem{21} K.J. Sankaran, J. Kurian, H.C. Chen, C.L. Dong, C.Y. Lee, N.H. Tai, and I.N. Lin, J. Phys. D \textbf{45}, 365303 (2012).

\bibitem{22} J.R. Arthur, J. Appl. Phys. \textbf{36}, 3221 (1965).

\bibitem{23} L.M. Baskin, O.I. Lvov, and G.N. Fursey, Phys. Status Solidi B \textbf{47}, 49 (1971).

\bibitem{24} D. Varshney, C. Venkateswara Rao, M. J. F. Guinel, Y. Ishikawa, B. R. Weiner, and G. Morell, J. Appl. Phys. \textbf{110}, 044324
(2011).

\bibitem{25} C. Ducati, E. Barborini, P. Piseri, P. Milani, and J. Robertson, J. Appl. Phys. \textbf{92}, 5482 (2002).

\bibitem{26} M. Liao, Z. Zhang, W. Wang, and K. Liao, J. Appl. Phys. \textbf{84}, 1081 (1998).

\bibitem{27} M. Cahay, P.T. Murray, T.C. Back, S. Fairchild, J. Boeckl, J. Bulmer, K.K.K. Koziol, G. Gruen, M. Sparkes, F. Orozco, and W. O'Neill,
Appl. Phys. Lett. \textbf{105}, 173107 (2014).

\bibitem{28} N.S. Xu, J. Chen, and S.Z. Deng, Appl. Phys. Lett. \textbf{76}, 2463 (2000).

\bibitem{29} J.P. Barbour, W.W. Dolan, J.K. Trolan, E.E. Martin, and W.P. Dyke, Phys. Rev. \textbf{92}, 45 (1953).

\end{thebibliography}
\end{document}